\documentclass[preprint,3p,12pt]{elsarticle}

\usepackage{amssymb}
\usepackage[dvipsnames]{xcolor}
\usepackage{amsmath}
\usepackage{float} 
\usepackage{subcaption} 
\usepackage{hyperref}
\usepackage{xurl}
\usepackage{booktabs}
\usepackage{titlesec}
\usepackage{makecell}
\usepackage[table]{xcolor}
\usepackage{booktabs}
\usepackage{multirow}
\usepackage{tabularx}
\usepackage{array}

\usepackage{fancyhdr}

\fancypagestyle{preprint}{
    \fancyhf{}
    \fancyhead[C]{\small\textit{Preprint -- September 2026}}
    
}

\definecolor{best}{RGB}{146,208,80}
\definecolor{intermediate}{RGB}{255,242,153}
\definecolor{worst}{RGB}{244,177,131}

\newcolumntype{Y}{>{\centering\arraybackslash}X}

\usepackage{graphicx}

\titleformat{\subsection}
  {\bfseries\upshape}
  {\thesubsection}
  {1em}
  {}

\makeatletter
\def\ps@pprintTitle{%
    \let\@oddfoot\@empty
    \let\@evenfoot\@empty
}
\makeatother

\begin{document}


\begin{frontmatter}


\title{Planning electric bus systems with solar photovoltaic integration using open transit data: A case study of the Dakar BRT}

\author[add1]{\texorpdfstring{Jérémy Dumoulin\corref{cor1}}{Jérémy Dumoulin}}
\ead{jeremy.dumoulin@epfl.ch}
\author[add2,add3]{Cheikh Mouhamed Fadel Kebe}
\author[add4]{Babacar M. Ndiaye}
\author[add1]{Noémie Jeannin}
\author[add1]{Christophe Ballif}
\author[add1]{Nicolas Wyrsch}

\cortext[cor1]{Corresponding author.}
\address[add1]{Photovoltaics and thin film electronics laboratory (PV-LAB), École Polytechnique Fédérale de Lausanne (EPFL), Institute of Electrical and Microengineering (IEM), Neuchâtel, Switzerland}
\address[add2]{Laboratoire Eau, Energie, Environnement et Procédés Industriels (LE3PI), Ecole Supérieure Polytechnique, Cheikh Anta Diop University of Dakar, Senegal}
\address[add3]{Centre de Test des Systèmes Solaires (CT2S) of Dakar, Senegal}
\address[add4]{Laboratory of Mathematics of Decision and Numerical Analysis, Cheikh Anta Diop University of Dakar, Senegal}


\begin{abstract}
Electrification of urban bus systems is expected to accelerate due to its clear environmental benefits. However, effective electric bus deployment requires planning tools that can support decision-making on electrification strategies while balancing operational feasibility, costs, environmental benefits, and impacts on the local electricity grid. Such planning depends on the availability of detailed transit data, which remain scarce in many cities, particularly in developing countries. To address this gap, we present GTFS4EV, an open-source framework that uses publicly available General Transit Feed Specification (GTFS) data to simulate bus operations and evaluate electrification scenarios. The framework provides quantitative insights across multiple dimensions. It estimates the minimum onboard battery capacity required for each bus, alongside charging infrastructure needs, economic and environmental impacts, and the potential for solar photovoltaic integration. We apply the framework to the Dakar Bus Rapid Transit system, comparing three charging strategies (“Depot only,” “Terminal and depot,” and “Terminal only”) across different levels of PV capacities. Results show that the “Terminal and depot” strategy reduces the minimum onboard battery capacity from 285 to 57~kWh per bus, and the maximum charging load from 5.3 to 1.5~MW relative to the “Depot only” strategy. The “Terminal only” also provides benefits compared to the “Depot only”, although it requires larger onboard battery capacities (155~kWh). Moreover, combining opportunity charging with solar photovoltaics reduces the charging costs by up to 46\% relative to grid charging only. The study demonstrates how GTFS data can be leveraged to support electric bus planning and photovoltaic integration in data-scarce contexts.
\end{abstract}

\end{frontmatter}

\thispagestyle{preprint}



\section{Introduction} 

\subsection*{Context and motivation}

The electrification of buses represents a key opportunity to decarbonize road transport, which still accounts for approximately 15\% of global energy-related CO$_2$ emissions \cite{IPCC_2022}. Beyond their potential to reduce greenhouse gas emissions, battery electric buses (BEBs) offer several additional benefits, including improved air quality, lower noise, and reduced operating costs \cite{Holland2021, Tsoi2023, Ghotge2025, Noll2026}. Driven by these advantages, together with declining battery prices and favorable policy support, their adoption has accelerated worldwide, with global sales approaching 70,000 units in 2025 \cite{EVOutlook2026}.

This global shift is particularly evident in Europe, where electric buses already account for more than half of new urban bus sales \cite{ICCT2026}. The European Union has further reinforced this transition through a target requiring all new urban buses to be zero-emission vehicles by 2035 \cite{EuropeanCommission2024}. Meanwhile, momentum is also building across many developing countries, where public transport already forms the backbone of passenger mobility \cite{DecarbAfrica}, making BEBs a natural pathway for passenger transport decarbonization. Notable examples include Ethiopia's ambitious e-mobility strategy \cite{EthiopiaPlan}, which targets the complete electrification of newly introduced public transport vehicles, or India's plans to replace 800,000 diesel buses by BEBs \cite{IEA_India_CaseStudy_2024}. As electrification expands, however, it also brings new challenges for planning and operating electric bus fleets.

\begin{figure}[ht]
    \centering
    \includegraphics[width=0.75\textwidth]{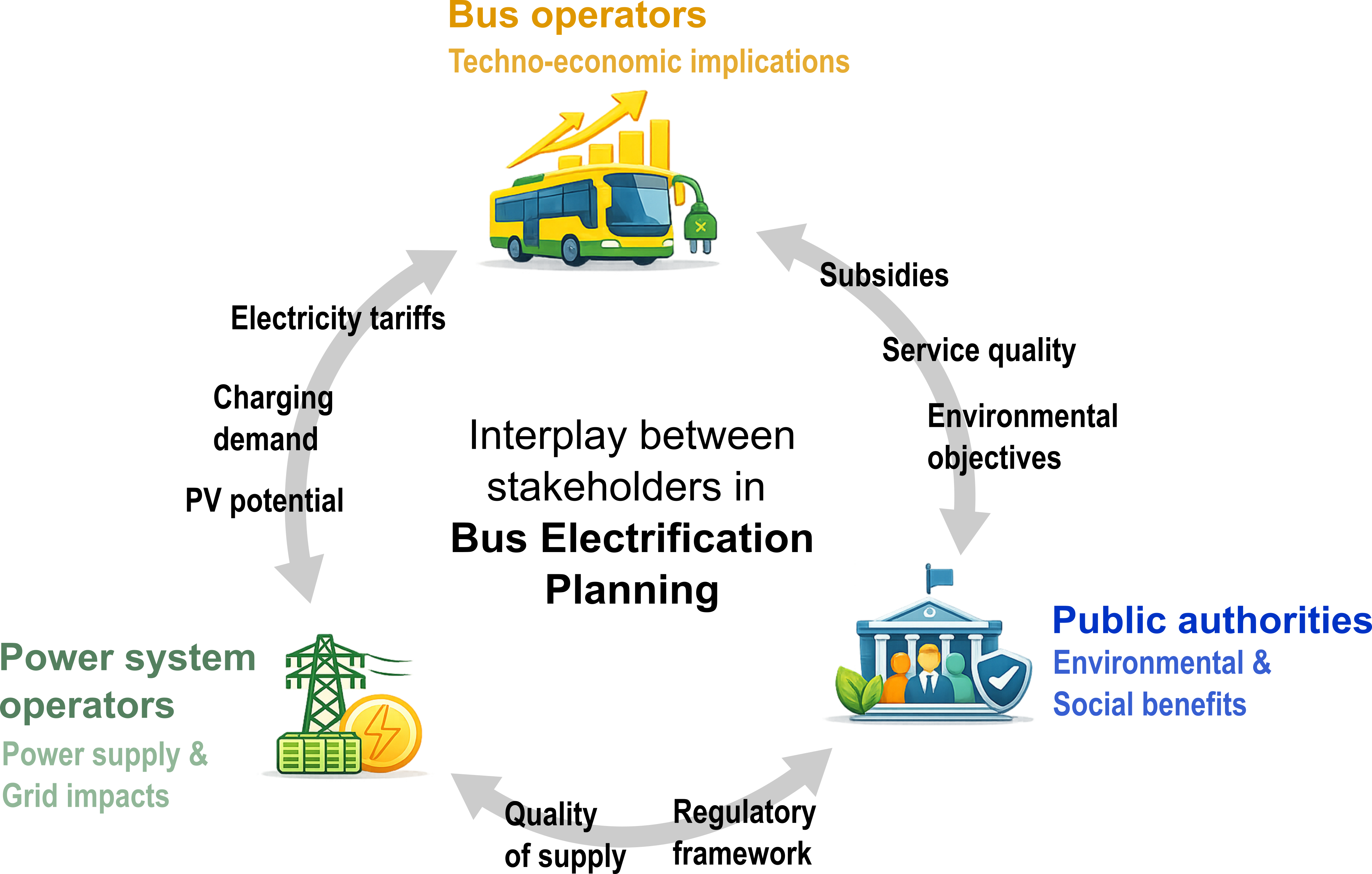}            
    \caption{Overview of the three main stakeholders involved in bus electrification planning, illustrating their key dimensions of interest, and mutual interactions.}
    \label{fig:stakeholder_interactions}
\end{figure}

As illustrated in Fig.~\ref{fig:stakeholder_interactions}, the transition to BEBs involves more than a simple vehicle substitution. It is a multi-stakeholder planning problem requiring coordinated decisions by bus operators, power system operators, and public authorities \cite{Haddad2026, Huang2026}. For bus operators, electrification introduces a range of techno-economic challenges, including decisions regarding onboard battery capacity, charging infrastructure deployment, and charging strategies, all while ensuring reliable operations \cite{Lee2021, Perumal2022}. For power system operators, the resulting charging demand must be accommodated without compromising grid stability \cite{Alamatsaz2022, Liu2024, Sharma2026}. From the perspective of public authorities, electrification must support broader environmental objectives while preserving the quality of transport services \cite{Perumal2022}. Consequently, planning tools are essential to navigate this complexity and to identify electrification pathways that are both technically feasible and aligned with diverse objectives.

The inherent complexity of this planning process is further intensified by a critical data gap. In particular, the analysis of bus electrification must rely on bus operation data that capture real-world travel behavior \cite{Perumal2022, Vijay2023}, including vehicle schedules, route structures, and dwell times at terminals and intermediate stops. Such detailed information is essential to estimate energy consumption and to define relevant charging strategies, which in turn determine key techno-economic requirements such as battery sizing and charging infrastructure \cite{Lee2021, Perumal2022, Heendeniya2023}. However, in many contexts, particularly in emerging economies, detailed vehicle-level data are not systematically collected by transit agencies or difficult to obtain \cite{Vijay2023, Booysen2025}. As a result, modeling studies must rely on assumptions or proxy data to represent bus operation, introducing uncertainty into planning outcomes and potentially resulting in suboptimal electrification pathways \cite{Heendeniya2023}.

\subsection*{Literature review}
A growing body of literature has focused on developing models and tools for BEB planning. Existing approaches span multiple planning stages, from strategic decisions on fleet size, vehicle specifications, and charging infrastructure to operational decisions on vehicle scheduling and charging \cite{Perumal2022}. Several recent reviews have synthesized these efforts \cite{Perumal2022, Alamatsaz2022, Behnia2024, Huang2026}, highlighting the diversity of methods used to represent vehicle operations, energy consumption, charging decision-making, and interactions with the charging and power infrastructures.

A key distinction in the literature concerns the level of detail used to represent bus operations. High-fidelity models generally rely on proprietary operational datasets, such as GPS trajectories, enabling very granular representations of vehicle operations \cite{Lee2021, Vijay2023}. However, these data are often unavailable outside well-instrumented transit systems and are rarely publicly accessible. At the opposite end of the spectrum, simplified screening tools based on generic operational assumptions (e.g., the electric vehicle charging tool released by the International Energy Agency \cite{IEA_EV_Charging_Grid_Tool_TechnicalNote_2023}) enable rapid first-order assessments but may overlook some operational characteristics that strongly influence the charging demand and investment decisions \cite{Perumal2022, Heendeniya2023}. Consequently, there is a need for open modeling approaches capable of leveraging openly available transit data while preserving sufficient operational realism to deliver context-specific insights.

Recent advances in the availability of open passenger-level transport data provide new opportunities to address this modeling gap. In particular, the General Transit Feed Specification (GTFS) has emerged as a widely adopted open standard \cite{gtfs_org}, providing harmonized data on routes, stops, timetables, and service frequencies. This data format is increasingly being made publicly available through transit agency websites and centralized repositories such as MobilityDatabase \cite{mobilitydatabase} or TransitLand \cite{transitland}. Furthermore, when not officially published, GTFS data can also be collected directly from passengers. Initiatives such as DigitalTransport4Africa \cite{DT4A} illustrate the feasibility of this approach, while also enabling data generation for informal transport modes such as minibus taxis \cite{Williams2015}, which lack a centralized operator able to generate GTFS data.

Despite these advances, GTFS-based frameworks for comprehensive BEB strategic planning remain limited. Existing studies mainly focus on reconstructing vehicle operations or estimating charging demand, while frameworks covering other technical dimensions of fleet electrification, multi-dimensional impacts, and renewable integration remain scarce. Importantly, as also highlighted by the recent review by Huang et al. \cite{Huang2026}, most studies only focus on economic impacts. To our knowledge, only Vijay et al. \cite{Vijay2023} proposed a GTFS-based framework covering multiple indicators, but it is primarily designed to provide high-level guidance for route electrification and does not consider the integration of renewable energy sources. In particular, the integration of distributed PV generation appears as a promising direction, as it can reduce operating costs, mitigate charging peaks, and improve the environmental performance of electric bus fleets \cite{Liu2024}. Moreover, the framework proposed by Vijay et al. \cite{Vijay2023} is not open source, limiting its transferability. The few existing open-source tools, such as EV-Fleet-Sim \cite{Abraham2021}, provide simulation capabilities for specific aspects of fleet electrification but do not address all requirements of strategic BEB planning.

Finally, despite the increasing diversity of modeling approaches, applications remain concentrated in Europe, North America, and selected regions of Asia. Evidence from Africa and other emerging economies remains limited, raising questions regarding the transferability of existing planning methodologies to contexts characterized by different mobility patterns, power grid constraints, and data availability \cite{Huang2026, Vijay2023}. These gaps highlight the need for open modeling approaches that can leverage openly available transit data while providing sufficient operational realism and considering the broader technical, economic, and environmental dimensions of BEB planning.

\subsection*{Identified gaps and contribution}

Overall, the literature review reveals four key gaps. First, existing open tools generally either rely on detailed vehicle-level datasets, such as GPS trajectories, or they adopt generic assumptions that enable rapid analyses at the expense of local realism. Hence, there is a need for open and transferable approaches capable of leveraging widely available datasets such as GTFS. Second, existing studies predominantly focus on economic criteria, whereas consideration including grid and environmental impacts, receive comparatively less attention. Third, the integration of distributed PV within bus electrification planning also remains limited, despite its potential to reduce operating costs and carbon emissions, and enable new business opportunities \cite{Liu2024}. Finally, existing applications are geographically concentrated in Europe, North America, and parts of Asia, raising questions about the transferability of current methodologies to emerging economies characterized by different objectives and constraints \cite{Huang2026, Vijay2023}. 

To address these gaps, this paper makes two main scientific contributions. The first is the development of GTFS4EV, an open-source strategic BEB planning framework that leverages GTFS data to simulate bus operation and evaluate electrification pathways. The framework combines operational simulation with assessments of the charging demand, the minimum onboard bus battery capacity for each bus, charging infrastructure requirements, environmental and economic implications, and the potential for solar PV integration. By relying on widely available transit data, GTFS4EV aims to support  decision-making for electric bus planning where detailed vehicle-level data are unavailable. 

The second contribution is application of GTFS4EV to the Dakar Bus Rapid Transit (BRT) system in Senegal, one of the first fully electric BRT systems in Sub-Saharan Africa. The case study evaluates alternative charging strategies and PV integration to quantify the trade-offs among battery sizing, charging infrastructure requirements, PV utilization, costs, and environmental performance. Specifically, this study addresses the following research questions:

\begin{itemize}
    \item To what extent can the Dakar BRT system be optimized through advanced on-route charging strategies while maintaining operational feasibility, and what trade-offs emerge among onboard battery capacity, charging demand, and charging infrastructure requirements?
    \item How can locally installed photovoltaic systems contribute to meeting the charging demand of the Dakar BRT under different charging strategies?
    \item What economic and environmental benefits can be achieved through the combined implementation of advanced charging strategies and photovoltaic integration?
\end{itemize}


\section{Methodology} 

\subsection*{Overview of the GTFS4EV framework}

\begin{figure}[ht]
    \centering
    \includegraphics[width=1.0\textwidth]{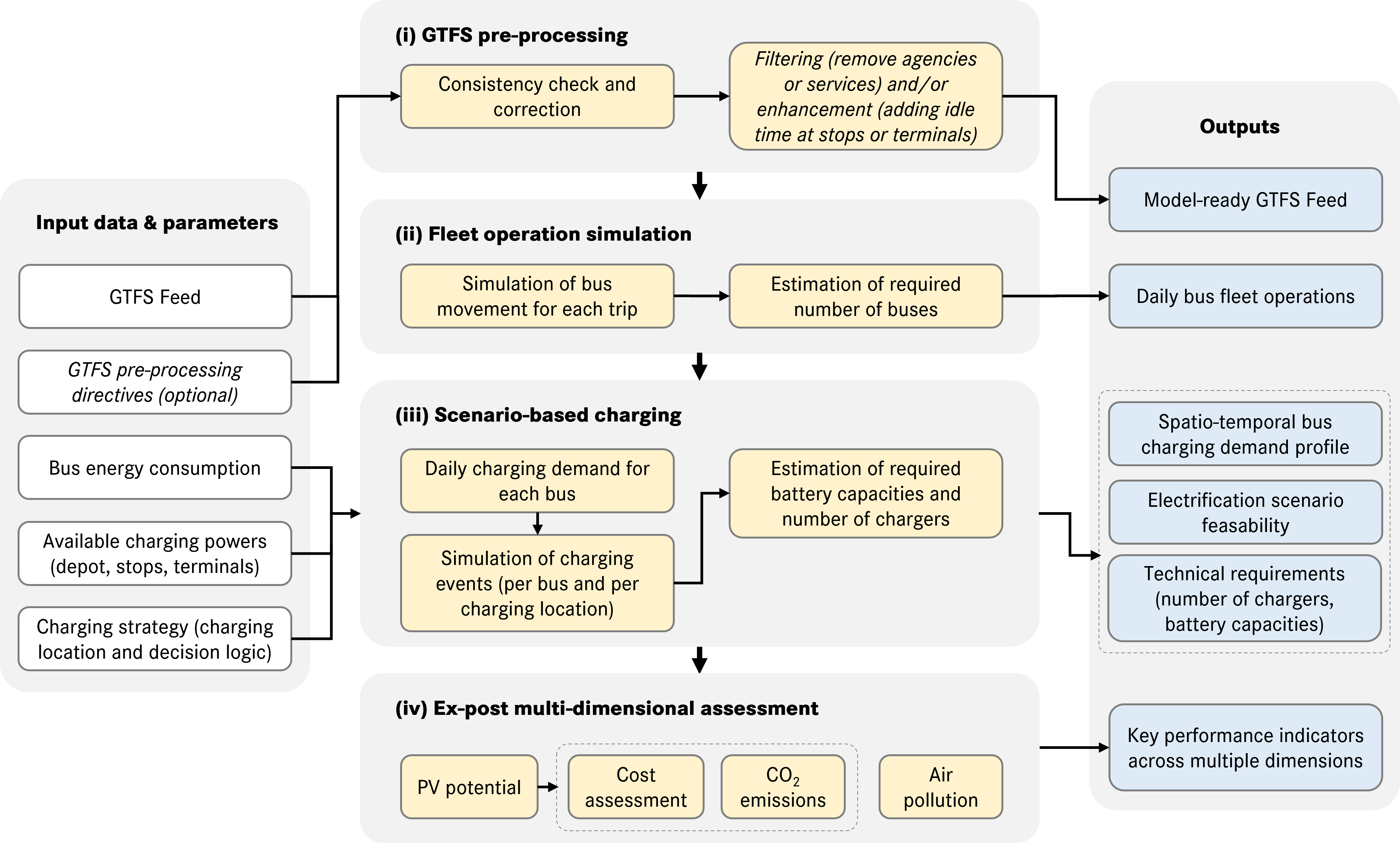}            
    \caption{Overview of the GTFS4EV framework and its workflow, showing the main inputs, outputs, and the four sequential modeling steps.}
    \label{fig:methodology}
\end{figure}

The methodology implemented in GTFS4EV consists of four main modeling steps, as illustrated in Fig.~\ref{fig:methodology}:
\begin{enumerate}
    \item GTFS data pre-processing: This first step ensures the internal coherence of the GTFS feed. Optionally, the GTFS feed can be filtered to restrict the analysis to a subset of routes or services and enriched with additional operational information, such as terminal dwell times.

    \item Fleet operation simulation: This step reconstructs vehicle operations by assigning buses to scheduled trips according to service frequencies, thereby determining the movement of each vehicle throughout the operating day. 

    \item Scenario-based charging: Charging events are simulated for each bus according to a user-defined electrification scenario. From these simulations, the scenario feasibility, spatio-temporal charging demand, and key technical indicators are derived (required onboard battery capacity and number of charging stations).

    \item Ex-post multi-dimensional assessment: Electrification scenarios are evaluated using a modular set of indicators, including PV integration potential together with economic and environmental performance metrics. 
\end{enumerate}

The framework is based on a modular architecture that enables rapid and flexible exploration of electrification pathways. While several predefined charging decision logics are implemented, additional strategies can be incorporated without modifying the core structure (see the online documentation of the GTFS4EV model \cite{gtfs4ev_docs}). In the present implementation illustrated in Fig.~\ref{fig:methodology}, key technical design variables, such as the required onboard battery capacity and the number of charging stations, are determined endogenously through the simulation of vehicle operations and charging events rather than being imposed exogenously. This approach is particularly valuable in early-stage planning, where fleet configurations and charging infrastructure have yet to be defined. Owing to its modular architecture, however, the framework could also accommodate exogenous design variables, enabling the evaluation of charging strategies under existing charging infrastructure or other operational constraints.

\subsection*{GTFS data pre-processing}

The first step validates and prepares the GTFS feed for simulation. GTFS4EV requires the following files: agency.txt, routes.txt, trips.txt, stop\_times.txt, stops.txt, calendar.txt, frequencies.txt, and shapes.txt. Although frequencies.txt is formally optional in the GTFS specification \cite{gtfs_org}, it is required by GTFS4EV because fleet reconstruction relies on headway-based service representations. A consistency check is performed to identify and remove invalid relationships between the required files, and the procedure is repeated iteratively until a coherent dataset is obtained.

Following this consistency pre-processing, optional filtering and enhancement procedures can be applied. Filtering allows analyses to be restricted to specific routes, operators, or service periods (e.g., weekday operations only). Data enhancement enables the incorporation of additional operational knowledge when GTFS schedules do not fully reflect observed operations. In particular, idle times at terminals and intermediate stops are sometimes omitted in GTFS feeds \cite{Mansurova2025}. GTFS4EV therefore allows users to specify additional dwell times based on local data or assumptions.

\subsection*{Fleet operation simulation}

The second step reconstructs vehicle operations from the pre-processed GTFS feed. Each GTFS trip is first represented individually before reconstructing the continuous daily operation of each vehicle. Here, a trip refers to a single scheduled bus run between its first and last stop, following the GTFS definition. For each trip, the vehicle trajectory is represented as a travel sequence, including sequences of traveling, stopping, and terminal dwell events. Route geometries are divided into segments between consecutive stops using the spatial information contained in stops.txt and shapes.txt. This event-based representation enables tracking the position of a vehicle over time, as well as the duration and distance associated with each segment.

Once the vehicle travel sequence has been established, the number of vehicles assigned to each trip is estimated from the service frequencies. For each trip, the number of vehicles required to operate the scheduled service is approximated as

\begin{equation}
N=\left\lceil\frac{T_{\mathrm{trip}}}{h}\right\rceil,
\label{eq}
\end{equation}

where $T_{\mathrm{trip}}$ denotes the duration of the trip, including any terminal dwell time associated with that trip, $h$ is the headway time, and $\lceil\cdot\rceil$ is the ceiling operator. Vehicles are dispatched along the trip at intervals equal to the headway time, thereby generating the operating schedule for each vehicle throughout the service period.

Since service frequencies may vary throughout the day, the corresponding fleet requirement also changes. Here, a conservative approach is adopted in which the number of vehicles assigned to a trip is fixed to the maximum number required across all service frequency periods. Consequently, some vehicles may remain idle during periods with lower service frequencies. Furthermore, GTFS feeds do not specify how vehicles enter (leave) service. In this work, the model assumes a transient start-up (shut-down) phase in which vehicles progressively enter (leave) service at the beginning (end) of the operating day.

\subsection*{Scenario-based charging}

The third step simulates the daily charging events of every bus according to a user-defined electrification scenario and derives the corresponding spatio-temporal charging demand profile, scenario feasibility, and associated technical requirements. Starting from the travel sequence obtained in the previous step, the cumulative energy consumed of a bus $b$ operating on a given trip is estimated from the sequence of travel events according to

\begin{equation}
E_{b}^{travel} = \sum_{i=1} e d_i + e d_{\mathrm{sec}}
\label{eq:energy_consumption}
\end{equation}

where $d_i$ is the distance traveled during segment $i$, $e$ is the average energy consumption (kWh.km$^{-1}$) over the service day, and $d_{\mathrm{sec}}$ is an additional reserve driving distance. This additional distance may represent a security margin or account for any additional bus movements which are not specified in the GTFS feed, such as trips to and from depots. 

Charging events are then simulated according to a charging strategy defining the available charging locations (depots, stops, or terminals) and a charging-decision logic. Whenever a vehicle reaches an eligible charging location, the charging conditions are evaluated and, if satisfied, the energy delivered during charging event $k$ is computed as

\begin{equation}
E_{b,k}^{charge} = \min\left(P_k\Delta t_k,E_{b,k}^{\mathrm{rem}}\right)
\label{eq:energy_charge}
\end{equation}

where $P_k$ is the charging power, $\Delta t_k$ is the charging duration, and $E_{b,k}^{\mathrm{rem}}$ is the remaining energy requirement before charging. This process is repeated throughout the service day to construct the charging schedule of each vehicle. For simplicity, no explicit charging efficiency is included in the formulation; however, the energy consumption $e$ can be readily adjusted to account for any charging losses.

\begin{figure}[ht]
    \centering
    \includegraphics[width=0.55\textwidth]{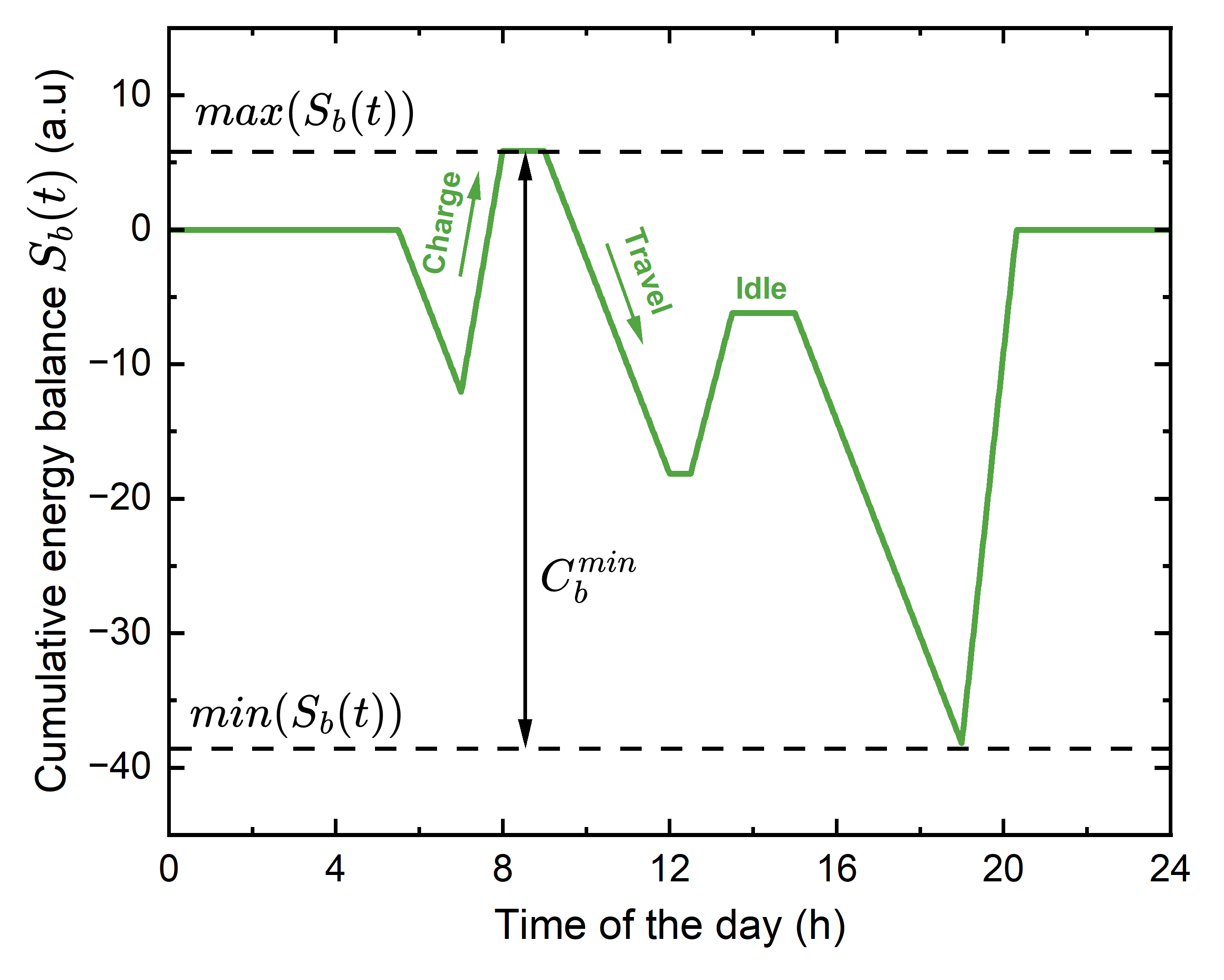}            
    \caption{Illustrative example of the cumulative energy balance $S_b(t)$ over a service day. Periods when the vehicle is traveling decrease the cumulative energy balance, while charging events increase it. The difference between the maximum and minimum values of the trajectory defines the minimum onboard battery capacity.}
    \label{fig:typical_energy_balance_evolution}
\end{figure}

The sequence of charging and travel events is also used to reconstruct the cumulative energy balance $S_b(t)$ of vehicle $b$ throughout the service day, corresponding to the cumulative charged energy minus the cumulative traction energy consumed, as illustrated in Fig. \ref{fig:typical_energy_balance_evolution}. From this, the minimum onboard battery capacity for a bus $b$ is obtained as

\begin{equation}
C^{\min}_b=\max\left(S_b(t)\right)-\min\left(S_b(t)\right),
\label{eq:battery_capacity}
\end{equation}

The proposed formulation assumes steady-state operation, whereby the state of charge at the beginning and end of the service day are equal. This eliminates the need for preliminary model runs to reach model convergence \cite{Pretorius2024}. In addition, the capacity given by Eq.\ref{eq:battery_capacity} represents the theoretical minimum required to complete the simulated service. The nominal capacity of an onboard battery must also account for other parameters like the allowable state-of-charge range, auxiliary consumption, and a reserve capacity. Unless otherwise stated, in the rest of the manuscript , "battery capacity" refers to this minimum onboard capacity for an individual bus.

Finally, by aggregating charging sessions by location, the minimum number of charging stations required at location $s$ can be derived from the maximum number of vehicles charging simultaneously

\begin{equation}
M^{\mathrm{ch}}_s=\max_t N_s(t),
\label{eq:number_chargers}
\end{equation}

where $N_s(t)$ denotes the number of vehicles charging at location $s$ at time $t$. When several trips share a same charging location, their charging schedules are aggregated prior to evaluating Eq.~\ref{eq:number_chargers}. Hence, the required charging infrastructure at a given location is determined from the peak simultaneous charging demand across all vehicles using that location.

\subsection*{Local PV production model}

Local PV production is simulated over one year at hourly resolution using the pvlib library \cite{Anderson2023}, combining local weather data from the PVGIS-SARAH3 database \cite{Jensen2023}, PV module specifications, and installation parameters. At each time step, the plane-of-array irradiance, $G_{\mathrm{POA}}(t)$ [W.m$^{-2}$], is computed from the irradiance data and PV module orientation. The electrical power output, $P_{\mathrm{PV}}(t)$ [W.m$^{-2}$], is then estimated using the \textit{PVWatts} model implemented in \textit{pvlib}

\begin{equation}
P_{\text{PV}}(t)=\eta_{\text{PV}}\,G_{\text{POA}}(t)\left[1+\beta\left(T_{\text{cell}}(t)-T_{\text{ref}}\right)\right]
\label{eq:pv_production}
\end{equation}

where $\eta_{\mathrm{PV}}$ is the nominal PV module efficiency, $\beta$ is the temperature coefficient [1/$^{\circ}$C], $T_{\mathrm{ref}}=25~^{\circ}$C, and $T_{\mathrm{cell}}$ is the cell operating temperature computed using the built-in \textit{PVsyst} thermal model. Angular losses are estimated using the formulation of Martin and Ruiz \cite{Martin2001}, while system losses are represented through an overall loss factor.

\subsection*{Multi-dimensional performance assessment}

The framework implements a modular set of indicators to evaluate electrification scenarios from different perspectives. Indicators can be selected according to the objectives of a given study. In the present work, the assessment focuses on fuel cost savings, CO$_2$ emission reductions, and the potential of locally installed PV to meet the bus charging demand. Other indicators supported by GTFS4EV, including local air pollutant emissions \cite{Dumoulin2026}, are beyond the scope of this work which mainly focuses on the additional benefits of PV integration in an already electrified bus system.

The local PV integration potential is characterized through three complementary indicators: the energy coverage factor, $CF$, the self-sufficiency, $SS$, and the self-consumption, $SC$ \cite{Simoiu2021}

\begin{equation} 
    CF = \frac{\int P_{PV}(t) dt}{\int P_{EV}(t) dt} 
    \label{eq:coverage_factor}
\end{equation}

\begin{equation} 
    SS = \frac{\int \min[ P_{PV}(t), P_{EV}(t) ] dt}{\int P_{EV}(t) dt}
    \label{eq:self_sufficiency}
\end{equation}

\begin{equation} 
    SC = \frac{\int \min[ P_{PV}(t), P_{EV}(t) ] dt}{\int P_{PV}(t) dt}
    \label{eq:self_consumption}
\end{equation}

where $P_{EV}(t)$ is the aggregated charging profile of the entire bus fleet at time $t$. While $CF$ quantifies the theoretical contribution of PV generation irrespective of temporal matching, $SS$ and $SC$ account for the coincidence between PV generation and charging demand and therefore characterize the effective use of PV electricity. The proposed metrics can be evaluated over different time horizons, from a single day to an entire year.

CO$_2$ emissions and fuel costs are both evaluated on an annual basis. The total CO$_2$ emissions associated with diesel and electric bus operation are estimated following

\begin{equation}
    CO_2^{diesel}=D_{year}c_{diesel}\gamma_{diesel}
    \label{eq:diesel_co2}
\end{equation}

\begin{equation}
    CO_2^{elec}=D_{year}e\gamma_{elec}
    \label{eq:electric_co2}
\end{equation}

where $D_{year}$ is the total annual distance driven. Diesel emissions are computed using $c_{diesel}$, the diesel fuel consumption (L.km$^{-1}$), and $\gamma_{diesel}$, the diesel CO$_2$ intensity (kgCO$_2$.L$^{-1}$). Electric vehicle emissions are computed from the electric energy consumption $e$ (kWh.km$^{-1}$) and the electricity carbon intensity $\gamma_{elec}$ (kgCO$_2$.kWh$^{-1}$).

Similarly, the total annual fuel costs associated with diesel and electric bus operation are computed as

\begin{equation}
    C_{diesel}=D_{year} c_{diesel}p_{diesel}
    \label{eq:diesel_cost}
\end{equation}

\begin{equation}
    C_{elec}=D_{year}e p_{elec}
    \label{eq:electric_cost}
\end{equation}

where $c_{\mathrm{diesel}}$ is the diesel consumption (L.km$^{-1}$), $p_{\mathrm{diesel}}$ is the diesel price (currency.L$^{-1}$), and $p_{\mathrm{elec}}$ is the electricity price (currency.kWh$^{-1}$).

When PV generation is considered, the electricity carbon intensity $\gamma_{elec}$ and the electricity price $p_{elec}$ are adjusted using the annual self-sufficiency and self-consumption 

\begin{equation}
    \gamma_{elec}=(1-SS)\gamma_{grid}+\frac{SS}{SC}\gamma_{PV}
    \label{eq:carbon_intensity}
\end{equation}

\begin{equation}
    p_{elec}=(1-SS)p_{grid}+\frac{SS}{SC} \, LCOE_{PV}
    \label{eq:effective_price}
\end{equation}

where $\gamma_{grid}$ and $\gamma_{PV}$ are the average carbon intensities of grid and PV electricity, respectively, $p_{grid}$ is the grid electricity tariff, and $LCOE_{PV}$ is the levelized cost of PV electricity generation. Such weighted-average formulations are widely used in the PV literature and provide a reasonable first-order quantification of PV integration benefits \cite{Vartiainen2024}. A detailed derivation of the effective electricity price is provided in \ref{app:electricity_price}. The same approach applies to the effective carbon intensity.


\section{Case study} 

\subsection{Urban mobility context in Dakar}

Dakar, located on the Cape Verde Peninsula in Senegal, is one of Africa's most densely populated metropolitan areas, concentrating nearly 4 million inhabitants within 547~km$^{2}$ \cite{ANSD2023RGPH5}. This represents almost 20\% of the national population on less than 0.3\% of the country's territory. This high density, combined with an unbalanced urban structure in which most employment opportunities are concentrated in the western part of the city while residential areas have expanded towards the east, has resulted in chronic traffic congestion and a range of negative externalities, including air pollution, noise exposure, and high commuting costs \cite{Lesteven2025, Setec2023Externalities}.

In this context, improving the local transport system has become a major priority for Dakar. Since the late 1990s, several initiatives have been launched to modernize the public transport sector. These efforts have included the construction of a toll highway, the Regional Express Train (TER) in 2021, and the deployment of a BRT system in 2024 \cite{Lesteven2025}. Building on these initiatives, the Dakar Urban Transport Executive Council (CETUD) recently adopted the ambitious \textit{Dakar Urban Mobility Plan 2035} \cite{PMUD}, which aims to develop an integrated, multimodal, and predominantly formal public transport system. The plan is structured around three complementary layers: a high-capacity backbone network based on mass transit systems (BRT and TER), a secondary network providing feeder services (regular buses), and a tertiary network ensuring last-mile connectivity. It also envisions the gradual phase out of informal transport modes \cite{SSATP2023, Lesteven2025}.

\subsection{The Dakar BRT}

\begin{figure}[ht!]
    \centering
    \includegraphics[width=0.6\textwidth]{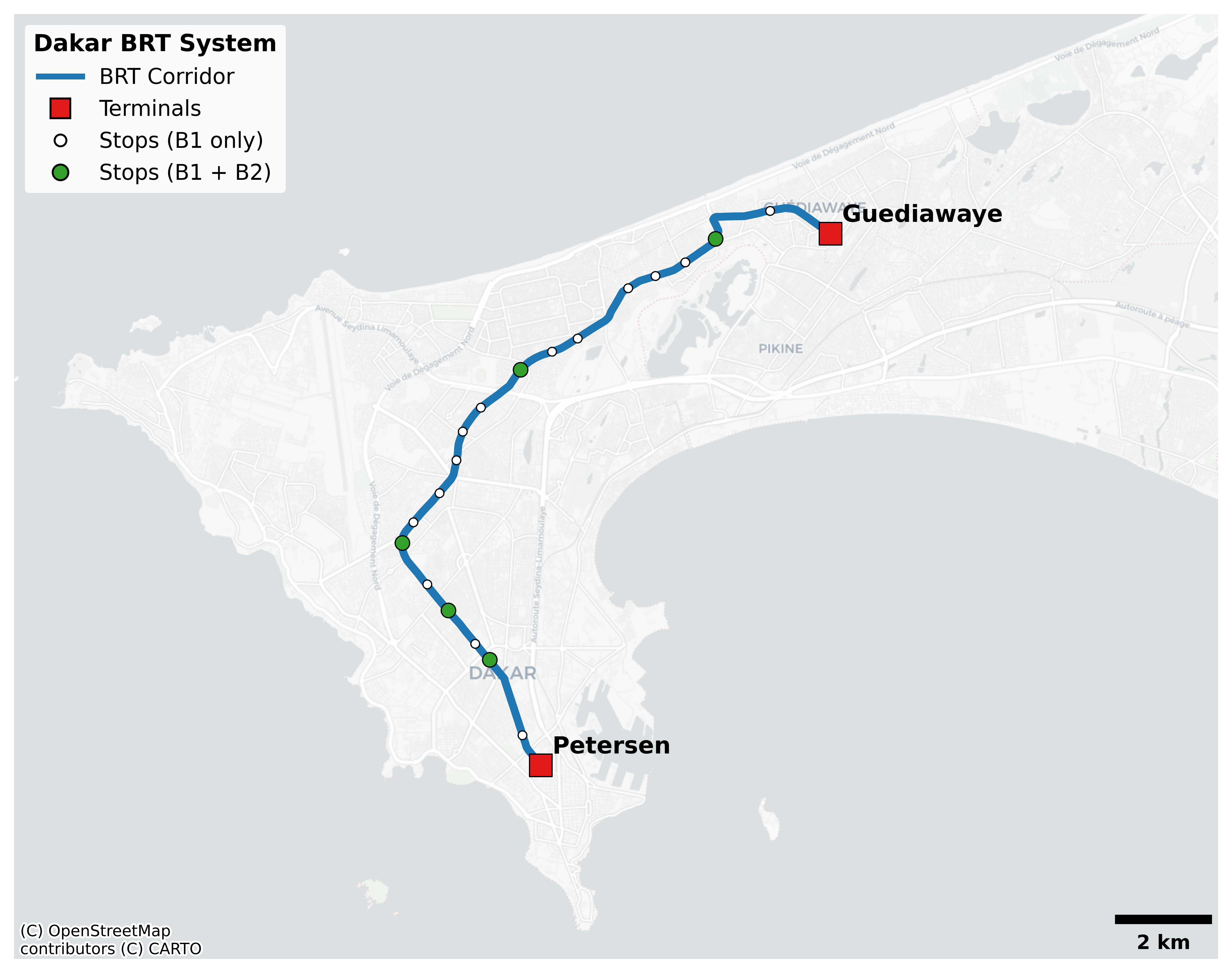}
   \caption{Map showing the BRT corridor between the Petersen and Guédiawaye terminals, as well as the intermediate 19 stops. The B2 "Semi-Express" line only serves the green stops, while the B1 "Omnibus" line serves all stops shown. Basemap source: CartoDB Positron.}
    \label{fig:dakar_brt_network}
\end{figure}

BRT is a bus-based transport system that combines dedicated lanes and transit signal priority to provide a high-capacity and reliable  transportation \cite{Cervero2013}. The Dakar BRT, progressively deployed since 2024, is an 18~km corridor designed to transport up to 300,000 passengers per day \cite{Taillandier2024DakarBRT,PMUD}. As depicted on figure~\ref{fig:dakar_brt_network}, it crosses the entire Dakar peninsula, connecting the Petersen bus terminal, located at the southern end of the corridor, to the Guédiawaye bus terminal in the north east. Once fully operational, the system will comprise four lines (B1 to B4), each serving a different number of stops. The entire bus fleet, which will ultimately comprise 121 buses, is fully electric.

At the time of this study, two lines were fully operational. Figure~\ref{fig:dakar_brt_network} also shows the stops of the Dakar BRT corridor. The first line, B1 "Omnibus", currently serves both terminals and 19 intermediate stops (i.e. 21 of the 23 stations planned in the final configuration are in operation). The second line, B2 "Semi-Express", serves five intermediate stops. Both lines operate from Monday to Saturday with a headway of 6 minutes between 6:00 and 21:00. On Sundays, only line B1 operates, with a 10 minute headway between 6:00 and 11:00  and a 7 minute headway between 11:00 and 21:00.

This BRT project represents a major step forward in Dakar's urban mobility plan, positioning the city as a pioneer in public transport electrification in Africa. However, there is still room to improve its environmental sustainability and reduce operational costs. Notably, the current charging strategy relies on overnight charging at bus depots, leaving the potential for daytime opportunity charging and PV integration untapped.
Exploring alternative charging strategies could therefore offer insights to and improve the current BRT system while informing the design of future electric BRT projects.

\subsection{GTFS data and model parameters}

\subsubsection*{GTFS feed and pre-processing}

The GTFS feed describes the transport schedules for the B1 and B2 lines for all days of the week. This study focuses exclusively on weekdays, as they represent a high-demand case in which service frequencies, and consequently charging demand and technical requirements, are at their highest. The provided GTFS feed adequately captures these weekday operating patterns, with minor pre-processing required to ensure consistency with current operating conditions (see~\ref{app:gtfs_preprocessing}). The analysis of the processed GTFS data identified three distinct bus service patterns considered in this study: the regular B1 "Omnibus" and B2 "Semi-Express", but also reinforcement buses operating on the B1 line during morning and evening peak travel hours.

\subsubsection*{PV system}

\begin{figure}[htbp!]
    \centering
    \begin{minipage}[b]{1.0\textwidth} 
        \centering
        \begin{subfigure}[t]{0.50\textwidth}
            \centering
            \caption{}
            \includegraphics[width=\textwidth]{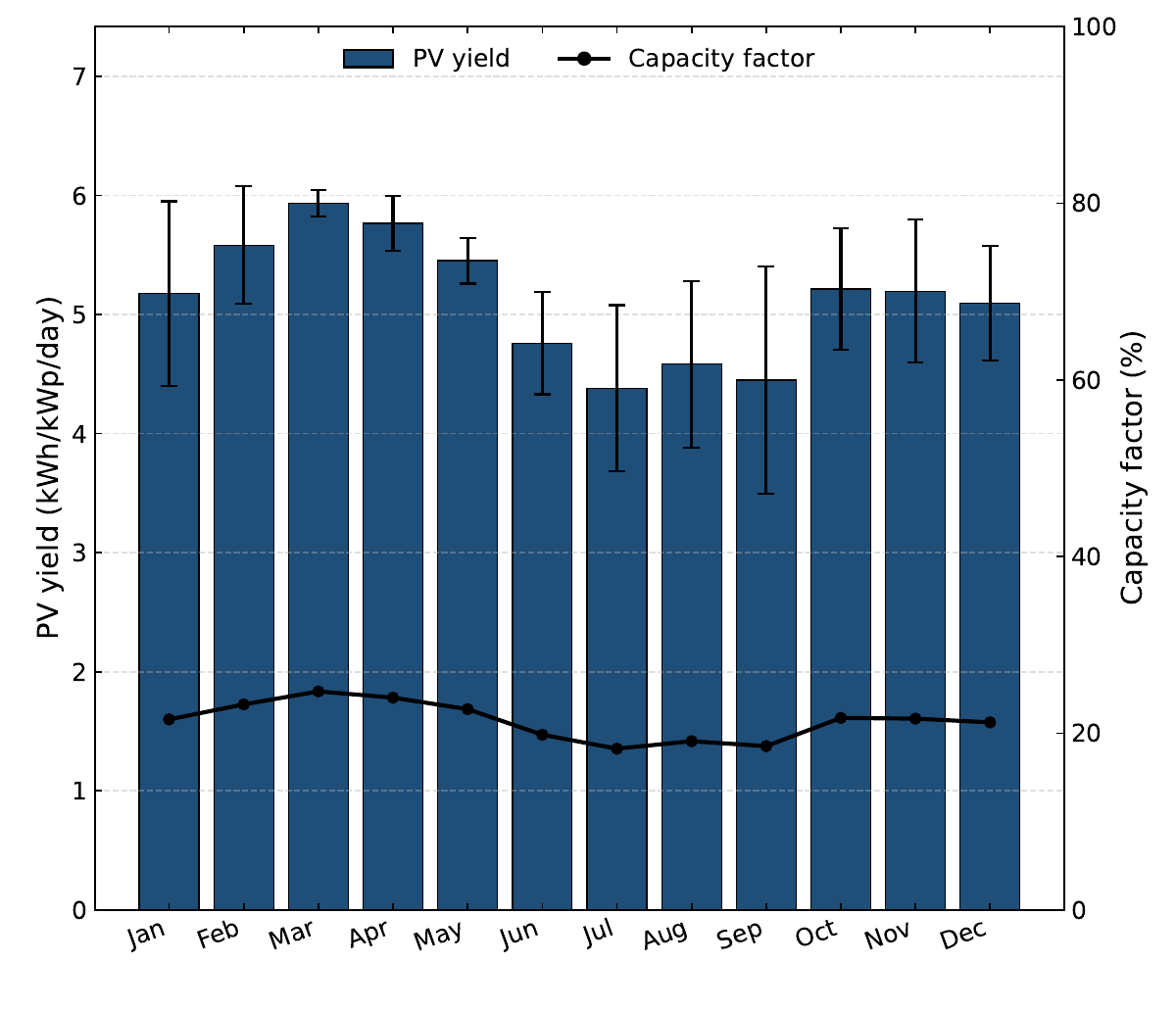}            
        \end{subfigure}
        \begin{subfigure}[t]{0.48\textwidth}
            \centering
            \caption{}
            \includegraphics[width=\textwidth]{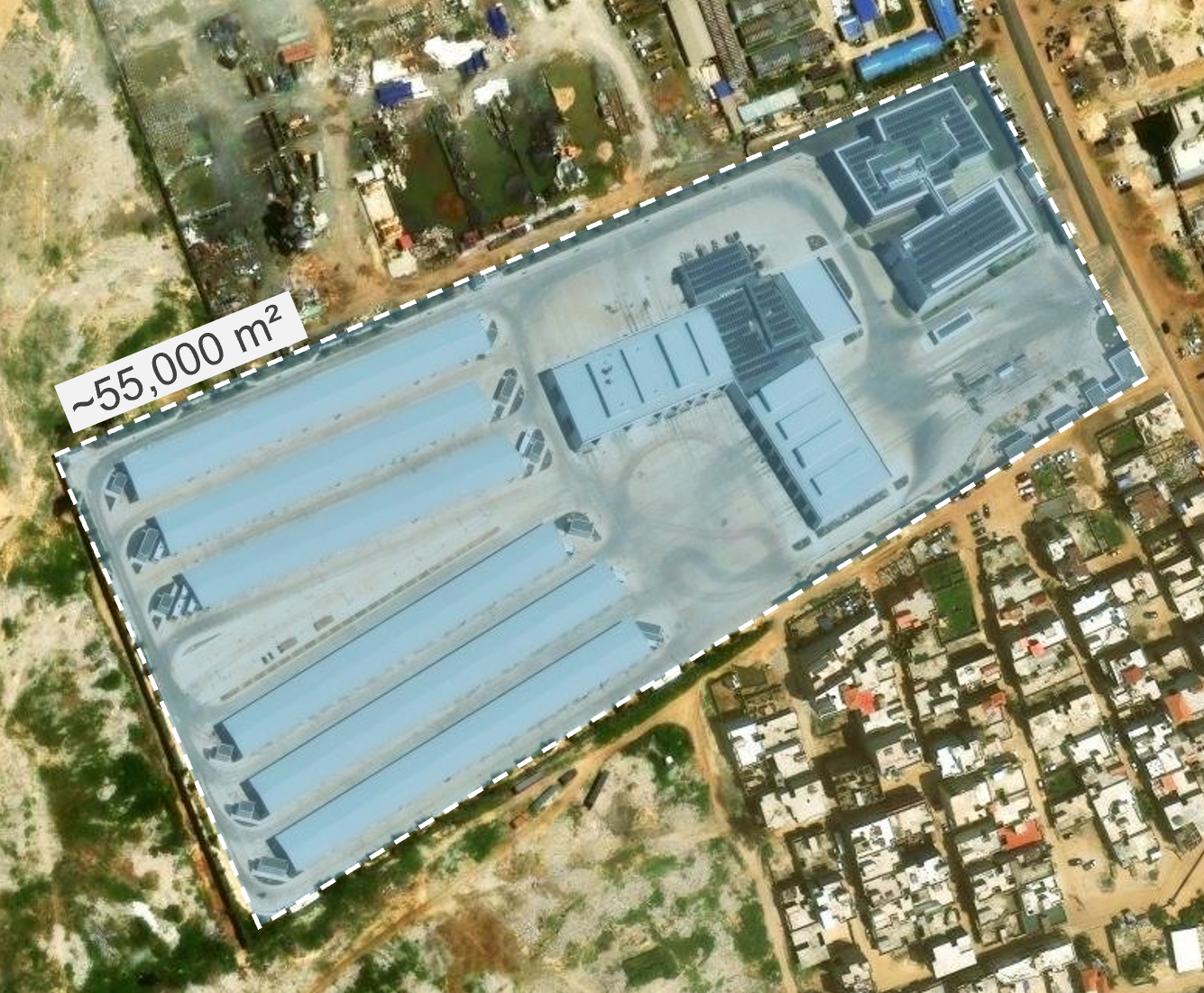}            
        \end{subfigure}
    \end{minipage}
    \caption{PV potential of the case study. (a) Monthly variation of the simulated daily PV energy yield and corresponding capacity factor. The bars represent the average daily PV yield with error bars indicating the monthly standard deviation, while the black line denotes the average capacity factor. (b) Satellite image of the bus depot in Gadaye, highlighting the available area for PV deployment (about 55,000~m$^2$).}
    \label{fig:pv_potential}
\end{figure}

This study considers free-standing PV systems installed at the optimal tilt angle, representative of PV installations on bus stations or canopy PV systems that could be deployed at bus depots or terminals. Weather data for Dakar for the year 2020 were used in the analysis. The PV system was modeled assuming a nominal module efficiency of 22\%, consistent with the recent sales-weighted average for commercially available PV modules \cite{FraunhoferISE_PVReport}. To reflect real-world operating conditions, a temperature coefficient of -0.4\%/$^{\circ}$C \cite{Jan2017} and overall system losses of 14\% were included in the simulation. The model predicts an optimal tilt angle of 17$^{\circ}$ and an annual energy yield of 1877.9~kWh/kWp. The lowest energy production occurs during the rainy season, which extends from June to September, as illustrated in Fig.~\ref{fig:pv_potential}a.

Based on the surface area available at key BRT facilities, the available PV deployment potential is estimated at 10~MWp. This first-order estimate includes the major bus depot located in Gadaye (see Fig.~\ref{fig:pv_potential}b), with approximately 55,000~m$^2$ of space available, the 23 planned bus stations, providing an additional 5,100~m$^2$, and the Grand Médine parking area, where some buses are parked during the day and where 10,000~m$^2$ of canopy-PV could easily be installed. Assuming a module efficiency of 22\% and a coverage factor of 60\% at the Gadaye depot, these areas correspond to approximately 10 MWp of installed capacity. This estimate is conservative: it excludes potentially usable space at bus terminals, other facilities owned by the bus operator, and future BRT infrastructures.

\subsubsection*{Fuel costs and CO$_2$ emissions}

The economic and environmental assessment relies on a set of reference parameters representative of current operating conditions in Dakar. A diesel consumption of 0.41~L.km$^{-1}$ \cite{Giraldo2019} and a diesel price of 1.184~USD.L$^{-1}$ \cite{FuelPrices} are assumed for conventional buses, while the electricity tariff was set to 0.20~USD.kWh$^{-1}$ during off-peak hours and 0.32~USD.kWh$^{-1}$ during on-peak hours (19:00–23:00) \cite{CRSE2026Tariffs}. 
For each charging scenario, the grid electricity price ($p_{grid}$ in Eq.~\ref{eq:effective_price}) is calculated by weighting the off-peak and on-peak rates by the charging energy in each period, using the simulated charging load curve. The grid electricity carbon intensity of Senegal is taken as 500~gCO$_2$.kWh$^{-1}$ and the diesel emission factor as 2.7~kgCO$_2$.L$^{-1}$ \cite{Dumoulin2026}.

For PV-based charging, we apply the methodology of Virtuani et al.~\cite{Virtuani2023} to Dakar to estimate the PV carbon intensity, resulting in 22.1~gCO$_2$.kWh$^{-1}$. This value is significantly lower than the average value for Europe (36~gCO$_2$.kWh$^{-1}$) reported in the reference study due to the higher solar resource in Dakar. As for the LCOE of PV generation, we estimate it as

\begin{equation}
LCOE_{PV} =
\frac{
C_{\mathrm{inv}} +
\sum_{t=1}^{N}
\frac{C_{\mathrm{O\&M},t}}{(1+r)^t}
}{
\sum_{t=1}^{N}
\frac{E_{PV,t}}{(1+r)^t}
}
\label{eq:lcoe_pv}
\end{equation}

where $C_{\mathrm{inv}}$ is the initial investment cost, $C_{\mathrm{O\&M},t}$ is the annual operation and maintenance cost in year $t$, $E_{PV,t}$ is the annual PV electricity production, $N$ is the project lifetime, and $r$ is the discount rate. The PV system is characterized by an investment cost of 1093~USD.kWp$^{-1}$, corresponding to the African average reported by \cite{IRENA2025}, and annual operation and maintenance costs of 10~USD.kWp$^{-1}$.yr$^{-1}$ \cite{IRENA2025}. A project lifetime of 25 years, a discount rate of 9\% \cite{IEA2024Senegal}, and an annual PV degradation rate of 0.95\%.yr$^{-1}$ \cite{Straub-Muck2025} are assumed. Based on these relatively conservative assumptions, the resulting PV LCOE is estimated at 0.0695~USD.kWh$^{-1}$. 

\subsection{Charging strategies}
At the time of this study, all buses are charged once a day at the depot outside operating hours using 120~kW chargers. A key objective of this work is to assess the potential benefits of alternative charging pathways, particularly with respect to increasing the utilization of on-site PV generation. As an initial step toward more advanced charging, three charging strategies are explored:

\begin{itemize}
    \item \textbf{Depot only}: Buses are charged exclusively overnight at the depot using 120~kW chargers. This baseline scenario represents the current situation.    
    \item \textbf{Terminal and depot}: Buses are charged both at terminals during layovers and at the depot overnight. Charging power is set to 120~kW at both locations, providing a scenario that is directly comparable to the baseline in terms of charging infrastructure. Whenever a bus arrives at a terminal, it charges at full power for the entire available layover duration.    
    \item \textbf{Terminal only}: Buses rely exclusively on 360~kW fast charging at terminals, a charging power representative of commercially available fast chargers for electric buses (e.g., \cite{KiiraMotors2026}).
    As a first-order charging-decision logic, a charging probability is introduced to distribute the charging demand throughout the day. This probability is set to 0.7 following the calibration procedure described in~\ref{app:charging_probability}, a value that spreads charging demand effectively while ensuring 100\% operational feasibility (i.e., all buses complete their required trips while satisfying their energy demand). This scenario evaluates the feasibility of fully daytime charging and its potential for direct utilization of local PV energy. 
\end{itemize}

Two major assumptions are made throughout this analysis. First, no bus rescheduling is considered: charging is restricted to existing layover periods and therefore does not affect service quality or timetable adherence. Second, only uncontrolled charging is considered, meaning that buses charge immediately at the maximum available power whenever they are plugged in, without any smart charging to improve, for example, PV self-consumption. Consequently, the considered scenarios represent a simple approach that requires minimal changes for bus operators and drivers. In addition, the energy consumption of the buses is assumed to be 1.3~kWh.km$^{-1}$, the value provided by the local transport authority. 


\section{Results} 

\subsection{Bus operation and the potential for alternative charging strategies}
Before simulating charging strategies alternative to the “Depot only” case, the operational characteristics of the bus fleet are analyzed using the results from Step 2 of the modeling framework. This analysis provides an initial screening of daytime charging opportunities by identifying when and where vehicles are idle and how operational patterns differ among the three services represented in the GTFS data: the B1 line, the B2 line, and the B1 reinforcement buses.

\begin{figure}[ht!]
    \centering
    \includegraphics[width=1.0\textwidth]{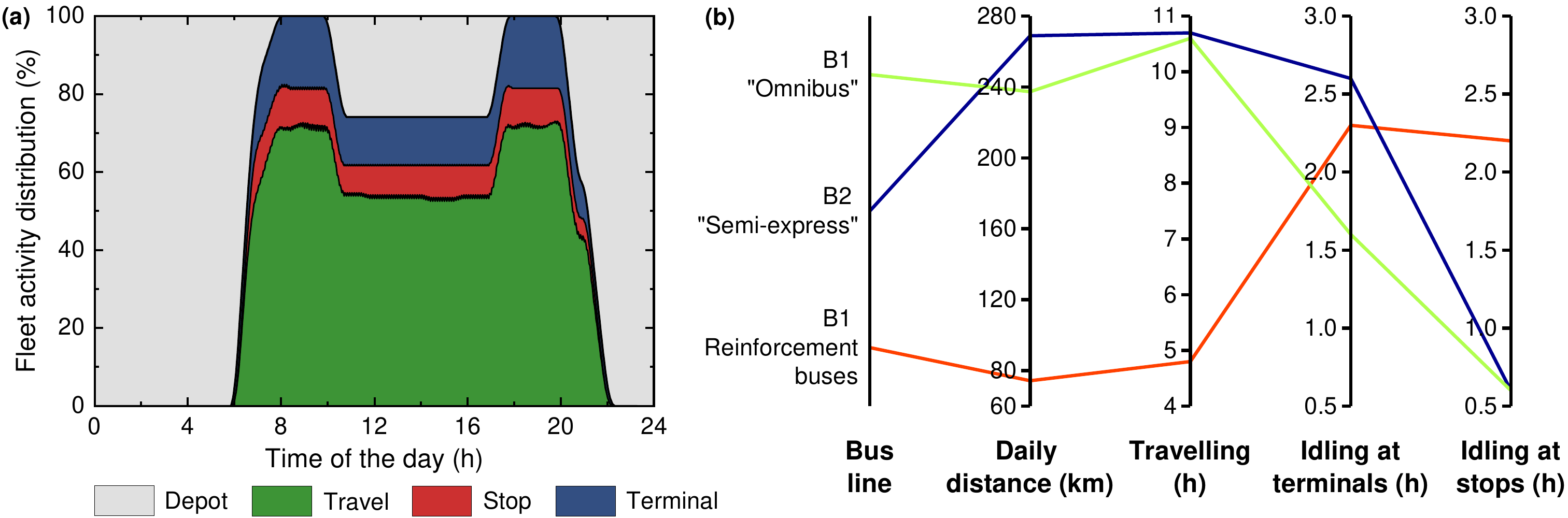}
   \caption{Temporal evolution of bus fleet activity over the day (panel a.). The stacked area plot shows the proportion of vehicles that are at the depot (gray), traveling (green), at stops (red) or at terminals (blue). A 10-minute moving average was applied to reduce short-term fluctuations. Panel b. shows differences between the three service patterns, showing the average per-bus daily distance traveled and the time spent traveling, and idling at terminals or stops.}
    \label{fig:fleet_activity}
\end{figure}

Figure~\ref{fig:fleet_activity}a shows the temporal evolution of bus fleet activity throughout a typical operating day. The fleet follows a clear daily operating cycle characterized by two peaks in the morning and evening, corresponding to the deployment of the reinforcement buses operating the B1 line between Grand Médine and Petersen during peak travel hours. As expected, the fleet deployment is progressive, with vehicles gradually entering service from 6:00 and returning to the depot from approximately 21:00. Traveling constitutes the dominant operating state, accounting for approximately 54\% of the fleet during off-peak travel hours and up to 72\% during the morning and evening peak travel hours. Nevertheless, a non-negligible share of vehicles remains stationary, revealing potential opportunities for daytime charging. Between 20\% and 28\% of the bus fleet is idling at stops or terminals during off-peak and peak travel hours, respectively. Most of this idle time occurs at terminals rather than intermediate stops, indicating that terminal layovers provide the greatest potential for opportunity charging. In addition, 26\% of the bus fleet remains at the depot during off-peak travel hours. These vehicles correspond to the B1 reinforcement buses, and could therefore also be charged at the depot between service periods.

Figure~\ref{fig:fleet_activity}b further characterizes the three service patterns by comparing their average per-bus daily distance and time spent traveling and idling. As expected, the B1 reinforcement buses exhibit the lowest operational intensity, traveling 74~km over 4.8~h per day because they operate exclusively during peak hours. In contrast, the B1 and B2 buses travel 237~km and 269~km, respectively, over similar durations (10.6~h and 10.7~h). The shorter distance covered by B1 buses reflects their more frequent stops. Despite these differences, all service patterns have long idle periods that could support opportunity charging: 1.6-2.6~h at terminals and 0.6-2.2~h at stops. Although B2 buses spend considerably less time at intermediate stops than B1 buses (0.6~h versus 2.2~h), their terminal idle times are comparable (2.6~h versus 2.3~h). These results identify terminal layovers as a charging opportunity across all service patterns, supporting the alternative charging strategies considered in this study.

\subsection{Charging load curve}

\begin{figure}[ht!]
    \centering
    \includegraphics[width=0.55\textwidth]{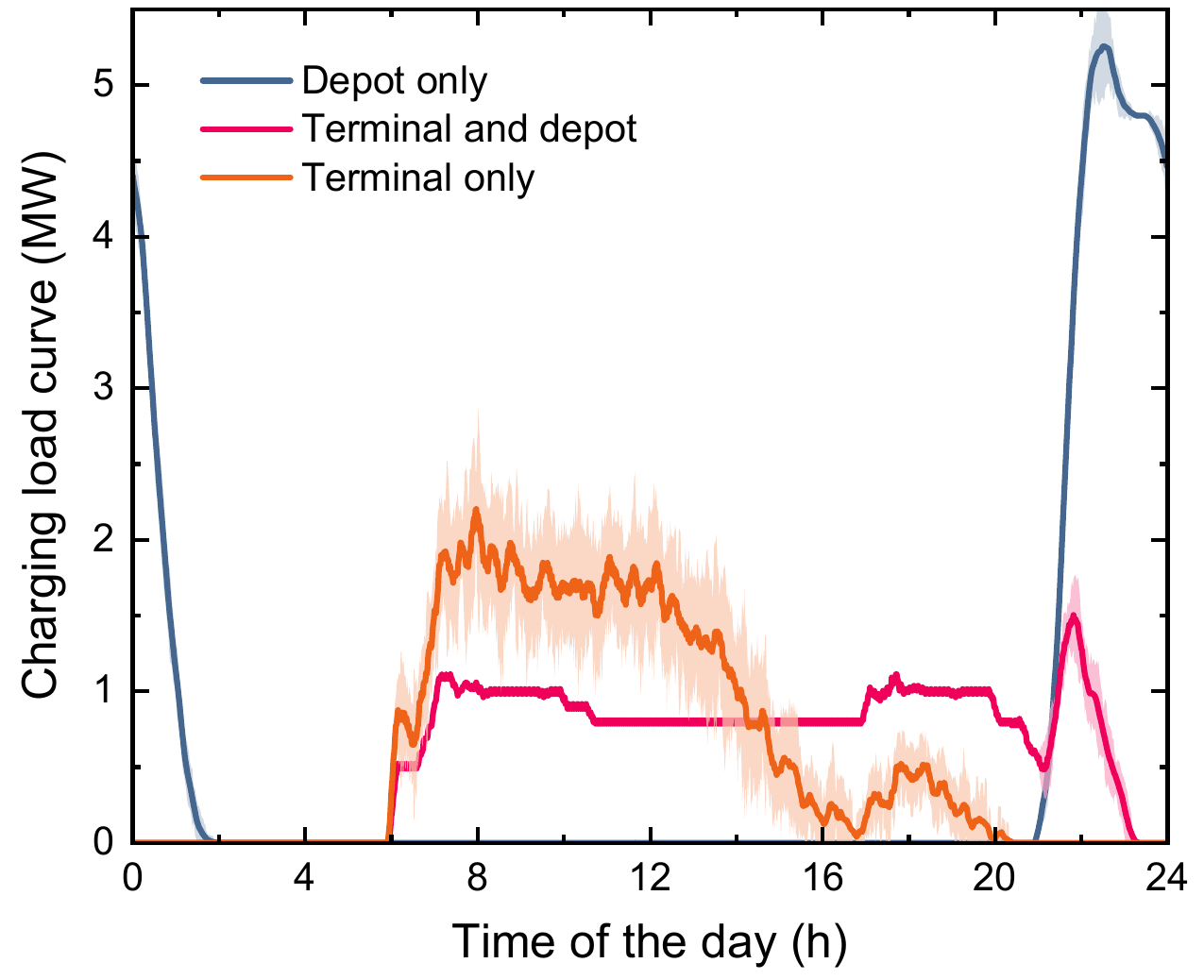}
   \caption{Charging load curves for the three charging strategies, aggregated across all charging locations and vehicles. For each scenario, the mean charging load (solid lines) and standard deviation (shaded areas) are reported over ten simulation runs in order to capture day-to-day variability arising from the stochastic model features.}
    \label{fig:charging_load_curve}
\end{figure}

Building upon the simulated bus operation, the following sections present a comparative analysis of the three charging strategies defined in Section~3. In this section, the impact of the charging strategy on the temporal distribution of the charging demand is analyzed through the aggregated charging load curve obtained from Step 3 of the modeling framework. As illustrated in Fig.~\ref{fig:charging_load_curve}, the charging strategy strongly affects both the shape and magnitude of the load curve.

The "Depot only" scenario leads to a pronounced overnight peak of 5.3$\pm$0.3~MW at 22:30, due to the simultaneous charging of most vehicles when returning to the depot at the end of the day. The "Terminal and depot" scenario reduces the overnight peak to only 1.5$\pm$0.2~MW. The remaining charging demand is spread throughout the day, resulting in a relatively flat daytime charging profile that remains around 1~MW. In addition, the overnight peak occurs around 40~minutes earlier than in the "Depot only" scenario, as vehicles arrive at the depot with lower remaining charging needs, allowing more vehicles to complete charging earlier. In the “Terminal only” scenario, charging demand occurs entirely during daytime operation. It remains at approximately 1.8~MW between 8:00 and 13:00, then declines because most buses have charged enough to complete their journeys. The greater fluctuations in this strategy result from the stochastic charging logic applied at terminals. Overall, these results show the potential of terminal-based charging to reduce the overnight peak at the depot by shifting the charging demand to daytime periods.

It is worth noting that the signature of the different service patterns presented in Fig.\ref{fig:fleet_activity}b are visible in the charging load curves of all scenarios. In the "Depot only" scenario, the early shoulder of the charging load curve is driven by the B1 reinforcement buses, which have small charging needs due to their short daily travel distances and therefore complete charging rapidly. The broader main peak is associated with the B1 and B2 buses, which have similar daily travel distances, resulting in greater charging needs. In the "Terminal and depot" scenario, two small increases in terminal charging demand can be observed during the morning and evening peak travel hours, from 07:00 to 10:00 and from 17:00 to 20:00, respectively. Similarly, in the "Terminal only" scenario, the charging demand occurring during the evening peak hours is primarily driven by the B1 reinforcement buses. In fact, the B1 and B2 buses require little or no additional charging during this period, as their energy demand has mostly been met through earlier terminal charging events.

\subsection{Battery and charging infrastructure requirements}
\begin{figure}[htbp!]
    \centering
    \begin{minipage}[b]{1.0\textwidth} 
        \centering
        
        \begin{subfigure}[t]{0.48\textwidth}
            \centering
            \caption{}
            \includegraphics[width=\textwidth]{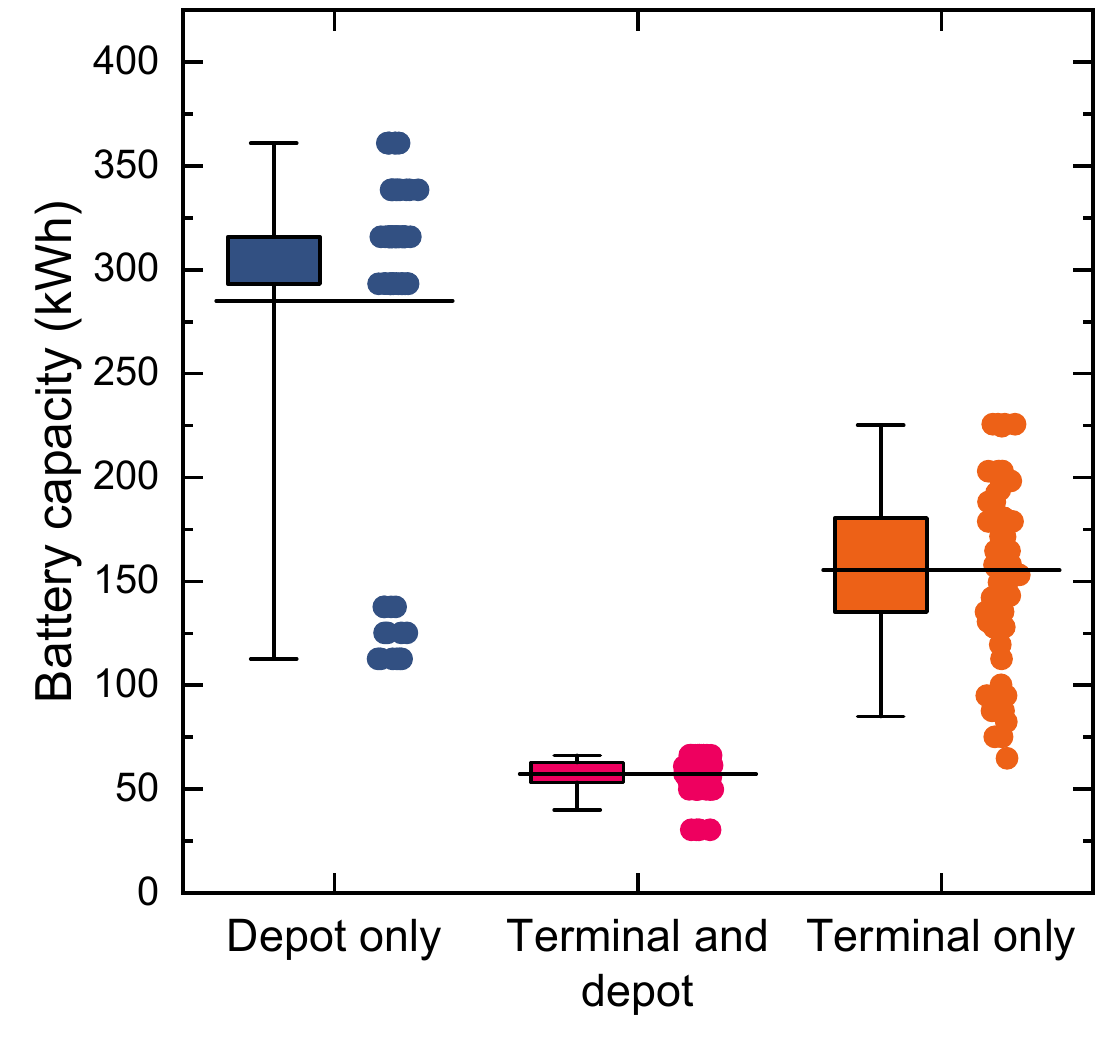}        
        \end{subfigure}
        \hfill
        \begin{subfigure}[t]{0.465\textwidth}
            \centering
            \caption{}
            \includegraphics[width=\textwidth]{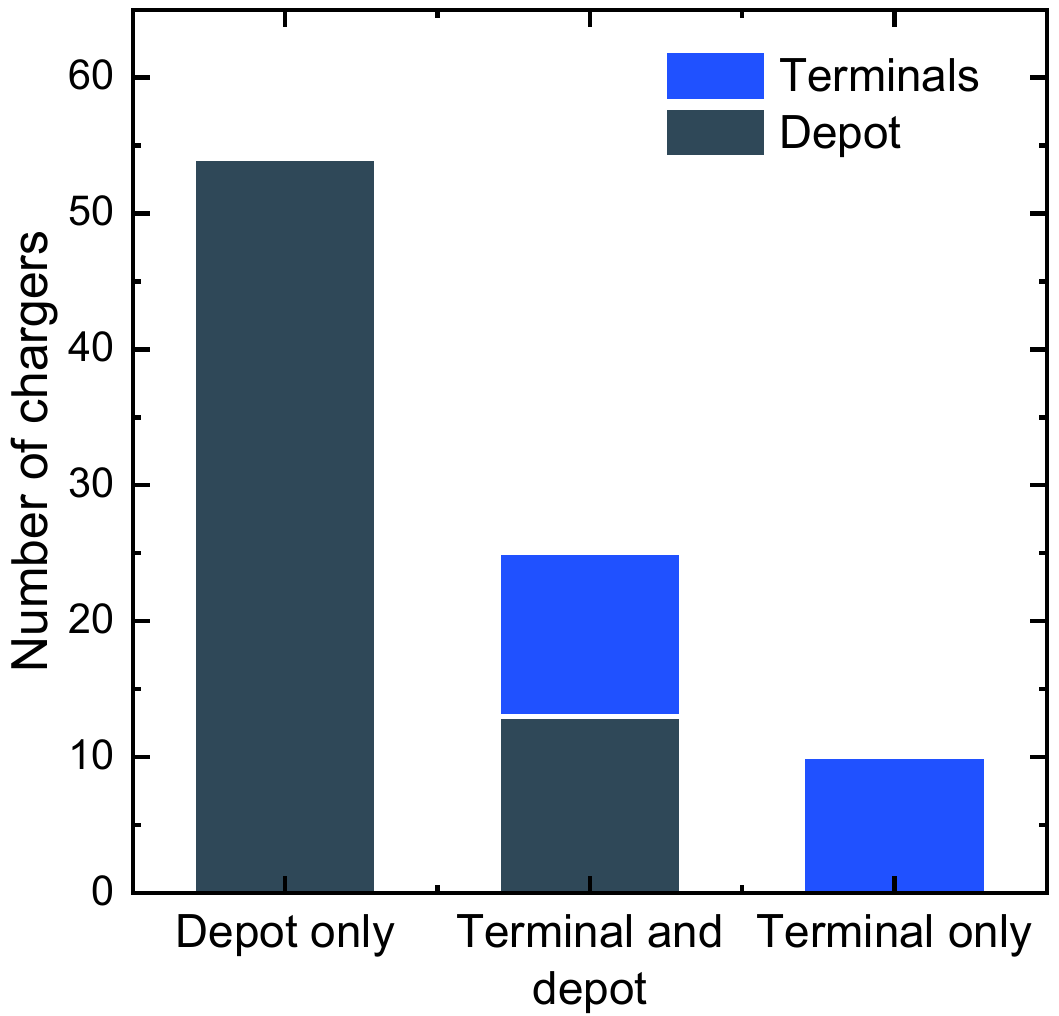}            
        \end{subfigure}
        
    \end{minipage}
    \caption{Impact of charging strategy on battery capacity and charging infrastructure requirements. (a) Minimum battery capacity required for each bus under the three strategies. Individual vehicles are shown as points, while box plots summarize the distributions (whiskers indicate the 5th and 95th percentiles, and the horizontal line shows the average).(b) Number of chargers required at depots and terminals for each charging strategy. }
    \label{fig:battery_and_chargers}
\end{figure}

Figure \ref{fig:battery_and_chargers}a shows the distribution of the minimum battery capacity required for the different buses to complete their daily service. The raw data points are shown for all three charging strategies and are accompanied by box plots, where the whiskers represent the 5th and 95th percentiles and the horizontal line indicates the average value. 

The "Depot only" strategy results in the highest battery capacities, with an average of 285~kWh. With this strategy, the battery capacity corresponds directly to the daily charging demand, because buses charge only once per day. Depending on the service pattern, the battery capacity thus varies considerably, reaching a maximum of 360~kWh. This maximum remains below the currently installed 540~kWh battery capacity, supporting the realism of the baseline charging strategy. In contrast, the "Terminal and depot" strategy results in the lowest required battery capacity, with an average value of only 57~kWh, due to the frequent charging opportunities distributed throughout the day. The “Terminal only” strategy represents an intermediate case, with an average battery capacity of 155~kWh. Because the high charging power may allow some buses to complete charging early in the day, they therefore require sufficient battery capacity to complete all remaining trips without further charging.

A closer examination of the battery capacity distribution shows distinct clusters, particularly visible in the "Depot only" strategy. Two levels of clustering can be distinguished. First, two broad clusters (approximately 110–140~kWh and 290–360~kWh in the "Depot only" strategy) arise because reinforcement buses have a substantially lower charging demand than buses operating on the other two lines. Within each broad cluster, a finer discrete structure is also visible, with many buses sharing identical battery capacities. This reflects differences in the energy demand of the B1 and B2 lines, together with small variations in the number of trips completed by individual buses. In fact, since some buses enter service later, they complete one or two fewer trips, leading to a discrete reductions in their daily energy demand. In the "Terminal and depot" and "Terminal only" scenarios, this clustered structure is less pronounced. In such opportunity charging strategies, the required battery capacity is largely determined by the charging frequency, which is similar across service patterns. Additionally, in the "Terminal only" strategy, the finer discrete structure is further smoothed out due to the probabilistic nature of this charging strategy.

The charging strategy also influences the required number of chargers (Fig. \ref{fig:battery_and_chargers}b). In the "Depot only" strategy, the number of chargers equals the fleet size, as each bus is assigned a dedicated charger at the depot. By enabling charger sharing through opportunity charging at terminals, the number of chargers is reduced in the other two cases. In the "Terminal and depot" strategy, the total number of chargers is cut by more than half, requiring only 13 chargers at the depot and 12 chargers at terminals. The "Terminal only" strategy further reduces the requirement to 10 chargers. This additional reduction is enabled by the higher charging power, which allows less frequent charging and, consequently, increases charger utilization. It should be noted that some variability exists in this case. However, additional simulations showed that this variation is small, with the required number ranging from only 9 to 11~chargers. Overall, these results demonstrate the strong potential of terminal-based charging to reduce charging infrastructure requirements while maintaining operational feasibility.

\subsection{Potential for PV-based charging}

\begin{figure}[ht!]
    \centering
    \includegraphics[width=0.99\textwidth]{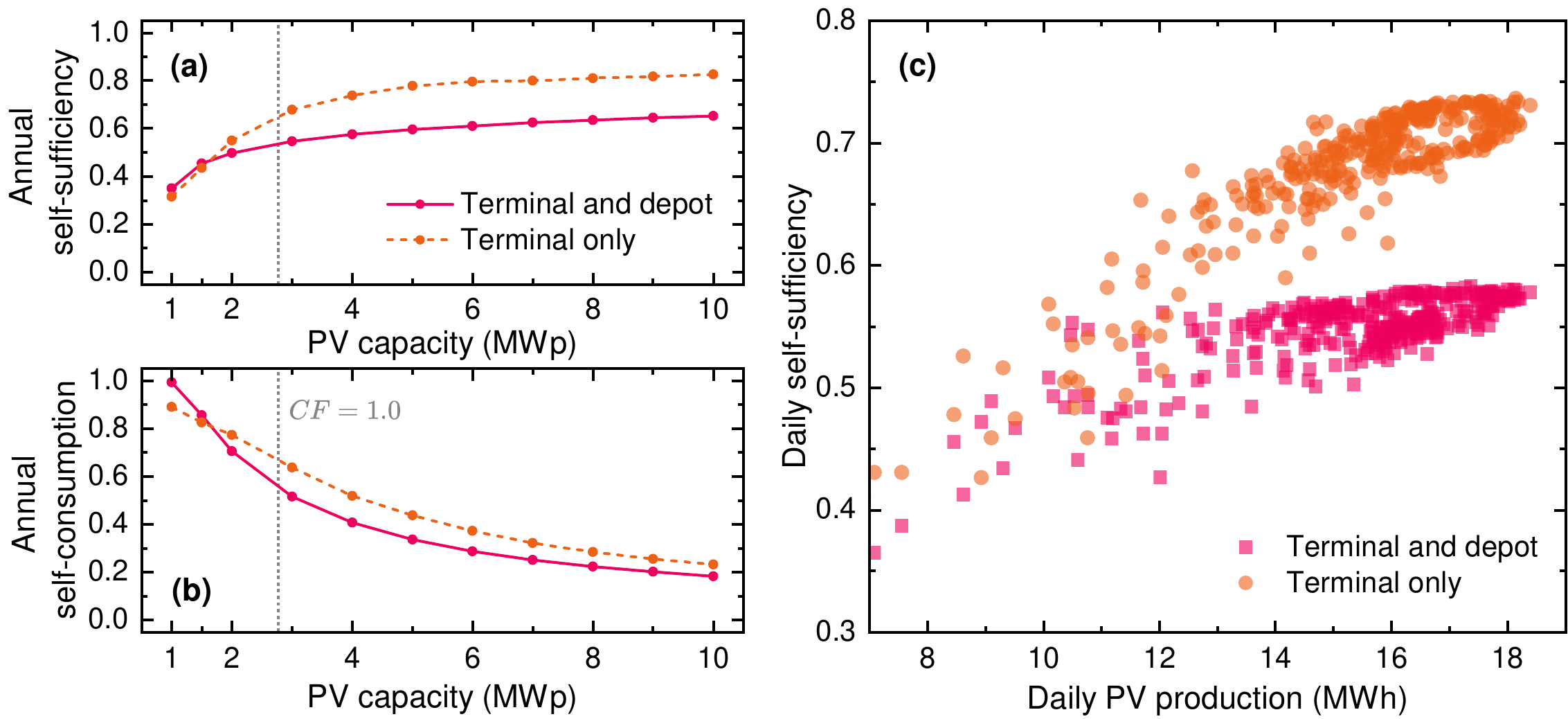}
    \caption{Effect of PV capacity and charging strategy on self-sufficiency and self-consumption. (a) Annual self-sufficiency and (b) annual self-consumption as a function of installed PV capacity for the "Terminal and depot" and "Terminal only" strategies. The vertical dashed line indicates the PV capacity at which the annual PV generation matches the annual charging demand ($CF=1.0$). (c) Daily self-sufficiency as a function of daily PV production for both strategies, considering all days of the simulated year.}
    \label{fig:ss_sc}
\end{figure}

The annual self-sufficiency and self-consumption, as defined in Section 2, are shown in Fig.~\ref{fig:ss_sc} for the two charging strategies that enable part of the bus fleet to be charged from PV during the day (i.e., "Terminal and depot" and "Terminal only"). The analysis is performed as a function of the total installed PV capacity, whose maximum potential for the BRT system is estimated at 10~MWp (see Section~3). The figure also indicates the PV capacity (2.8~MWp) at which the annual PV generation equals the annual charging demand ($CF=1.0$). The two strategies achieve high levels of annual self-sufficiency and self-consumption, with values exceeding 50\% at the 2.8~MWp threshold in both cases. If the full PV potential of 10~MWp is utilized, self-sufficiency increases to 65\% and 82\% for the "Terminal and depot" and "Terminal only" scenarios, respectively. These values are particularly noteworthy given that they are achieved using only direct PV charging, without any additional flexibility provided by stationary battery storage or smart charging.

The results also reveal a crossover between the two strategies at approximately 1.5~MWp. Below this PV capacity, the "Terminal and depot" scenario performs better, achieving a self-sufficiency between 35\% and 45\% while maintaining a self-consumption above 85\%. Its relatively flat charging demand profile allows PV energy to be utilized throughout the day as long as the PV production remains below the charging demand, resulting in a more efficient use of the available solar energy. However, beyond 1.5~MWp, an increasing share of the PV production, particularly around solar noon, can no longer be consumed directly, causing self-consumption to decrease and self-sufficiency to gradually plateau at approximately 65\%. In contrast, the "Terminal only" charging load profile better matches the midday PV production peak, allowing self-sufficiency to continue increasing up to 82\%, while maintaining also a higher self-consumption.

To examine the day-to-day variability, Fig.~\ref{fig:ss_sc}c shows the daily self-sufficiency against the daily PV production for the 2.8~MWp capacity over all days of the year. Daily self-sufficiency ranges from 23-58\% for the "Terminal and depot" strategy and from 27-74\% for the "Terminal only" strategy, with a clear monotonic relationship observed for both charging strategies, confirming that daily PV production is the primary driver of self-sufficiency. Overall, the "Terminal only" strategy generally achieves higher daily self-sufficiency for a given level of PV production, with its advantage becoming more pronounced as PV production increases. This is because its charging demand is better aligned with the midday peak in PV generation, allowing a larger fraction of the available solar energy to be consumed on sunny days. In contrast, the "Terminal and depot" strategy reaches this maximum more quickly. However, it also exhibits a slightly narrower range of daily self-sufficiency and even outperforms the "Terminal only" strategy on less sunny days. This highlights a trade-off between maximizing self-sufficiency on sunny days and reducing sensitivity to day-to-day weather variability.

\subsection{Implications on carbon emissions and fuel costs}

\begin{figure}[ht!]
    \centering
    \includegraphics[width=0.55\textwidth]{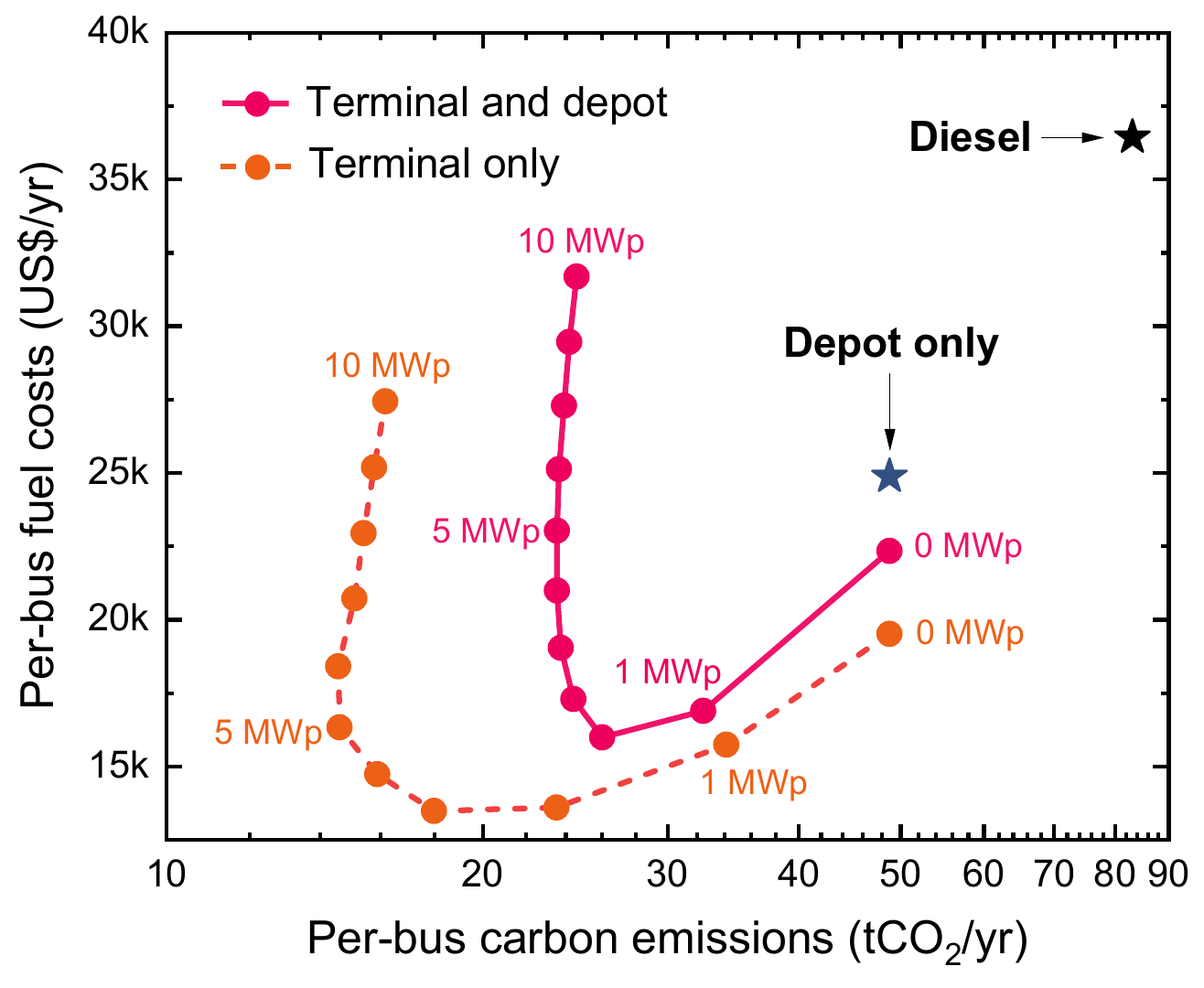}
    \caption{Annual average per-bus carbon emissions versus fuel costs for the "Terminal and depot" and "Terminal only" charging strategies under different installed PV capacities. Each point represents installed PV capacities ranging from 0~MWp to 10~MWp, in 1~MWp increments. The diesel case and the "Depot only" charging strategy are included for comparison.}
    \label{fig:pareto}
\end{figure}

The annual per-bus CO$_2$ emissions and fuel costs for the "Terminal and depot" and "Terminal only" charging strategies are presented as a function of the installed PV capacity in Fig.~\ref{fig:pareto}, alongside the fully grid-based “Depot only” strategy and the diesel case. The results confirm that electrification can substantially improve both environmental and economic performance under all charging strategies. Compared with the diesel baseline, annual per-bus CO$_2$ emissions decrease from 83.1~tCO$_2$.yr$^{-1}$ to 48.8~tCO$_2$.yr$^{-1}$ in the "Depot only" case. Similarly, the annual fuel cost is reduced from US\$36.4k to US\$24.9k, despite a large fraction (about 46\%) of the charging energy being supplied during the more expensive on-peak tariff periods in this charging strategy. The "Terminal and depot" and "Terminal only" strategies further reduce both CO$_2$ emissions and fuel costs over a wide range of installed PV capacities, with the lowest values achieved with the "Terminal only" case, highlighting the potential benefits using locally generated PV electricity.

Increasing PV capacity reduces the annual per-bus CO$_2$ emissions for both opportunity charging strategies, reaching a minimum at around 5~MWp. At this point, emissions fall to 23.5 and 14.6~tCO$_2$.yr$^{-1}$ for the “Terminal and depot” and “Terminal only” cases, respectively. At low PV capacities, CO$_2$ emissions decrease rapidly because nearly all additional PV generation is consumed by the fleet, directly replacing grid electricity. Beyond the 5~MWp threshold, emissions start to increase slightly as self-sufficiency plateaus (Fig.~\ref{fig:ss_sc}) and more PV electricity is exported to the grid. Thus, the embodied emissions of additional PV capacity are no longer fully compensated by reductions in grid imports. Nevertheless, emissions remain well below those of the “Depot only” strategy, confirming the environmental benefits of PV integration even beyond the optimal PV capacity.

Fuel costs appear to be more sensitive than CO$_2$ emissions to the installed PV capacity. Similar to the observed CO$_2$ emissions trend, fuel costs initially decrease sharply with increasing PV capacity, reaching a minimum at approximately 2~MWp for the “Terminal and depot” strategy and 3~MWp for the “Terminal only” strategy. These thresholds are lower than the 5~MWp associated with minimum CO$_2$ emissions, showing that the cost- and CO$_2$-optimal PV capacities do not coincide. Nevertheless, fuel costs remain below those of the “Depot only” strategy up to 6~MWp for “Terminal and depot” and 9~MWp for “Terminal only.” At the minimum, annual fuel costs fall to US\$15.9k and US\$13.5k, respectively, with the latter corresponding to approximately half the cost of the “Depot only” strategy. This reduction is primarily driven by direct PV charging, with additional savings from off-peak charging, as shown by the cost difference between the “Depot only” and opportunity charging strategies at 0~MWp.


\section{Discussion} 

From a methodological perspective, this study demonstrates how widely available GTFS data can support strategic planning for bus fleet electrification. The proposed open-source framework provides a practical compromise between high-fidelity models and simplified screening approaches, enabling the rapid exploration of alternative electrification scenarios with few assumptions. Notably, it demonstrates that a GTFS-based framework can capture differences in service patterns and provide sufficiently granular insights such as bus-specific battery capacity requirements, as illustrated in Fig.~\ref{fig:battery_and_chargers}. Its applicability also extends beyond planning the transition to electric buses, as illustrated by the Dakar case, where it supports the evaluation of alternative charging strategies for an already electrified fleet. 

The study also reveals a broader methodological outcome that extends beyond the Dakar case study: investment requirements are strongly shaped by the charging strategy, suggesting that strategic and tactical planning should be considered jointly rather than sequentially, as is traditionally done in bus planning \cite{Perumal2022}. In this sense, the charging strategy emerges as a key design variable rather than merely an operational decision. Recognizing this interdependence can lead to more cost-effective electrification strategies and calls for a more integrated bus planning paradigm.

By combining indicators across multiple dimensions, the GTFS4EV framework enables a multi-stakeholder assessment of electrification strategies and a quantitative exploration of trade-offs among the objectives of bus operators, power system operators, and public authorities. Table~\ref{tab:stakeholder} summarizes the main indicators and trade-offs across the charging strategies considered for the BRT case study. The "Depot only" strategy consistently performs poorly, confirming the potential of opportunity charging, while the comparison between the "Terminal and depot" and "Terminal only" strategies reveals trade-offs. For instance, although "Terminal only" favors PV integration, it requires greater battery capacity and results in higher peak charging power without PV. Note that detailed results on the effect of PV integration on peak power are not presented in this study; however, a complementary analysis shows that "Terminal only" with PV integration can reduce the aggregated peak charging demand, consistent with previous studies on solar-powered charging in Africa \cite{wevj17060313}.

\begin{table}[t]
\centering
\caption{Qualitative comparison of the charging strategies across the indicators considered in this study, grouped by the primary stakeholders interested in these indicators. For  indicators influenced by PV integration, the assessment is provided with (w/) and without (w/o) PV. Ratings are relative to the strategies evaluated in this paper.}
\label{tab:stakeholder}

\begin{tabularx}{\textwidth}{>{\raggedright\arraybackslash}p{2.2cm} >{\raggedright\arraybackslash}p{5.6cm} Y Y Y}
\toprule
\textbf{Stakeholder} & \textbf{Indicator} &
\textbf{Depot only} &
\textbf{Terminal and depot} &
\textbf{Terminal only} \\
\midrule

\multirow{3}{*}{\makecell[l]{Bus\\operator}}
& Number of charging stations
& \cellcolor{worst} Worst
& \cellcolor{intermediate} Medium
& \cellcolor{best} Best \\

& Battery capacity
& \cellcolor{worst} Worst
& \cellcolor{best} Best
& \cellcolor{intermediate} Medium \\

& Fuel costs (w/ and w/o PV)
& \cellcolor{worst} Worst
& \cellcolor{intermediate} Medium
& \cellcolor{best} Best \\

\midrule

\multirow{2}{*}{\makecell[l]{Power system\\operator}}
& Peak charging power (w/o PV)
& \cellcolor{worst} Worst
& \cellcolor{best} Best
& \cellcolor{intermediate} Medium \\

& Peak charging power (w/ PV)
& \cellcolor{worst} Worst
& \cellcolor{intermediate} Medium
& \cellcolor{best} Best \\

\midrule

\multirow{1}{*}{\makecell[l]{Public\\authority}}
& CO$_2$ emissions (w/ and w/o PV)
& \cellcolor{worst} Worst
& \cellcolor{intermediate} Medium
& \cellcolor{best} Best \\

\bottomrule
\end{tabularx}

\end{table}

Regarding PV integration, the results demonstrate a strong potential for both opportunity charging strategies. In terms of CO$_2$ emissions, they outperform the "Depot only" scenario across the entire range of PV capacities. Fuel costs can also be reduced by up to 36\% ("Terminal and depot") and 46\% ("Terminal only") relative to the "Depot only" strategy. Over a 12-year vehicle lifetime, this corresponds to cumulative fuel savings of approximately US\$108k ("Terminal and depot") and US\$137k ("Terminal only") per bus. Given that the purchase price of a BEB typically amounts to several hundred thousand dollars, depending on the bus size and battery capacity \cite{Sistig2025}, these savings represent a substantial contribution toward offsetting the higher upfront investment compared with diesel buses.

From an economic perspective, however, both the PV capacity and the charging strategy strongly influence the achievable fuel savings, with a small risk of oversizing the PV system within the estimated 10~MWp potential. Nevertheless, the economically attractive PV capacity range increases as the charging strategies improves, as illustrated by the "Terminal only" compared with "Terminal and depot" one. This suggests a practical deployment pathway in which PV capacity is expanded progressively as BEB charging becomes more sophisticated. Since the two opportunity charging strategies considered here are not directly designed to maximize PV utilization, and neither smart charging, stationary battery storage, nor feed-in tariffs were considered, the economic benefits reported here are  conservative, leaving room for further improvement. Notably, the considered charging strategies do not exploit the potential of daytime depot idling periods, representing a low-hanging opportunity to further improve PV integration.

Battery sizing also has important economic implications. Based on the average battery capacities reported Fig.~\ref{fig:battery_and_chargers}, and on a battery cost US\$350~kWh.$^{-1}$ \cite{Sistig2025}, the reduction relative to the “Depot only” scenario would correspond to US\$79,800 per bus for the “Terminal and depot” strategy and US\$45,500 per bus for the “Terminal only” strategy. These estimates reflect only first-order differences in energy capacity, not final battery sizing which must account for other effects such as operating margins, degradation, and auxiliary loads. 

The two opportunity charging strategies also reduce the required number of charging stations, although the implications for infrastructure costs are less direct. Industry feedback gathered for this study suggests that a 360~kW charger costs approximately twice as much as a 120~kW charger. The reduction in chargers (Fig.~\ref{fig:battery_and_chargers}) may therefore offset the higher unit cost. The comparison in Fig.~\ref{fig:battery_and_chargers} should nevertheless be read in light of the assumed charging sharing assumptions. In the "Depot only" strategy, is is assumed that each bus has a dedicated charger, so coordinated charger sharing could reduce the number of chargers. Similarly, at terminals, the model assumes that buses can charge without queuing; competition for chargers could affect both the number of chargers needed and the achievable charging schedules.

We acknowledge several limitations in this study. While the input parameters for the BRT case study were selected, as far as possible, in accordance with guidelines from the local transport authority, a few parameters were derived from informed assumptions (e.g., available space for PV integration). We also apply the same energy consumption rate to all buses, as provided by the local transport authority. This does not capture possible differences between lines arising, for example, from their stopping patterns \cite{Blades2024}. Nevertheless, because the B1 and B2 lines share the same corridor and therefore elevation profile, we expect these differences to be small. Should more information become available, the model could be further refined to reflect additional technical and economic details of the BRT system. Such refinements would also be required for a realistic total cost of ownership analysis, which would be the next step toward a full economic comparison of the strategies, beyond fuel costs alone. Regarding PV integration, GTFS4EV aggregates PV generation and charging demand across the entire system, effectively assuming that generation at one location can offset demand at another. This approach captures the overall potential of PV integration and its capacity to reduce upstream grid congestion but does not provide insight into the effects at individual charging sites. Such a detailed analysis is left for future work and will depend on the spatial distribution of PV generation and charging demand, as well as local grid constraints. Finally, our study is limited to the year 2020. However, this assumption is expected to have a limited impact on the results, as annual variations in solar irradiance are small (data from PVGIS-SARAH3 indicates year-to-year fluctuations of approximately 2\% in Dakar).


\section{Conclusion} 

This study has demonstrated how widely available GTFS data can support strategic planning for electric bus systems through the development and application of the open-source GTFS4EV framework. The comparative charging strategy analysis of the Dakar BRT shows that, compared with the baseline “Depot only” strategy, the “Terminal and depot” strategy reduces the average battery capacity from 285 to 57 kWh, lowers the peak charging demand from 5.3 to 1.5~MW, and cuts the number of required chargers by more than half. The “Terminal only” strategy further reduces the requirement in terms of number of chargers and better aligns the charging demand with daytime PV generation, although it requires larger battery capacities (155 kWh) and fast chargers to meet the bus charging demand exclusively through opportunity charging at terminals. Combining opportunity charging with solar PV reduces the annual charging costs by up to 46\% relative to grid-based charging while significantly lowering CO$_2$ emissions. More generally, our findings demonstrate that the charging strategy and the investment requirements (battery capacity, charging infrastructure, PV sizing) are closely coupled. For electric bus systems, strategic and tactical planning should therefore be considered jointly rather than sequentially, requiring a shift away from conventional bus planning. GTFS4EV paves the way for such an integrated planning approach while providing a transferable decision-support tool for regions where detailed vehicle-level operational data remain unavailable.


\appendix

\setcounter{figure}{0}
\renewcommand{\thesection}{Appendix \Alph{section}} 

\section{Derivation of the effective electricity price}
\label{app:electricity_price}

The annual electricity cost $C$ of the BRT system can written as the sum of grid electricity purchases and the cost of PV generation:

\begin{equation}
C = p_{\mathrm{grid}}(E_d - E_{PV,used}) + LCOE_{PV} E_{PV},
\end{equation}

where $E_d$ is the annual electricity demand for bus charging, $E_{PV,used}$ is the PV electricity directly used by the buses, $E_{PV}$ is the total PV electricity generated, and $LCOE_{PV}$ is the levelized cost of PV electricity.

Dividing by the annual demand $E_d$ yields the effective electricity price:

\begin{equation}
p_{elec} = \frac{C}{E_d} = p_{\mathrm{grid}}\left(1 - \frac{E_{PV,used}}{E_d}\right) + LCOE_{PV}\frac{E_{PV}}{E_d}.
\end{equation}

Introducing the self-sufficiency $SS$ and self-consumption $SC$, as defined in Section 2, and substituting into the expression of $p_{elec}$ therefore leads to

\begin{equation}
p_{elec} = (1-SS)p_{grid} + \frac{SS}{SC} LCOE_{PV}.
\end{equation}

It should be noted that $LCOE_{PV}$, $SS$, and $SC$ are not independent. Increasing PV capacity typically increases self-sufficiency while reducing self-consumption, and the economic viability of PV integration depends on the balance between these two effects. In particular, PV integration ceases to reduce the effective electricity cost when $\frac{LCOE_{PV}}{SC} > p_{grid}$.

Finally, this formulation does not include any value associated with excess PV electricity exported to the grid. Such feed-in remuneration could be incorporated as an additional revenue stream or, equivalently, interpreted as a mechanism that effectively reduces the net cost of electricity consumption. Its exclusion therefore corresponds to a conservative case where exported electricity is not compensated.

\section{Charging probability calibration}
\label{app:charging_probability}

For the "Terminal only" charging strategy, a simplified charging logic is implemented through the introduction of a charging probability, $p_{\mathrm{charge}}$. This parameter represents the probability that a bus initiates a charging event each time it arrives at a terminal. 

The objective of this approach is to distribute charging events throughout the day without requiring an explicit optimization algorithm or predictive charging control. A lower value of $p_{\mathrm{charge}}$ reduces the likelihood of simultaneous charging events, thereby spreading the charging demand over a longer period and potentially increasing the direct utilization of PV generation while reducing peak power demand. However, low values may lead to insufficient charging opportunities and therefore reduced feasability. Therefore, $p_{\mathrm{charge}}$ must be as low as possible to promote charging demand dispersion but sufficiently high to guarantee charging strategy feasibility.

\begin{figure}[ht!]
    \centering
    \includegraphics[width=0.55\textwidth]{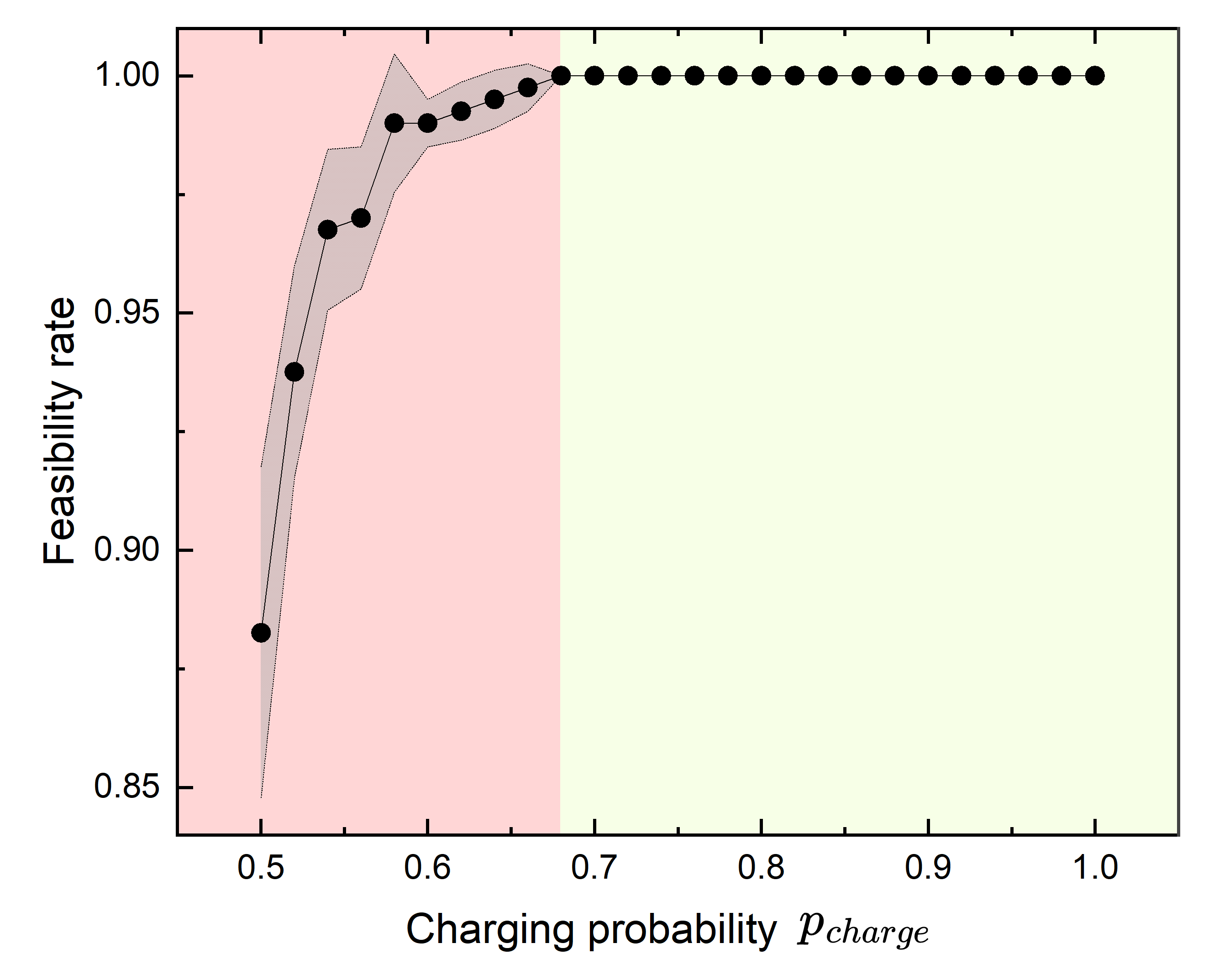}
    \caption{Feasibility rate as a function of the charging probability $p_{\mathrm{charge}}$ for the \textit{Terminal only} scenario. The envelope represents variability across 5 model runs. The selected value ($p_{\mathrm{charge}}=0.7$) ensures 100\% feasibility rate.}
    \label{fig:charging_probability}
\end{figure}

To identify an appropriate value, a parametric calibration procedure was performed by varying $p_{\mathrm{charge}}$ between 0.5 and 1.0. The resulting operational feasibility is presented in Fig. \ref{fig:charging_probability}. The results show that 100\% feasibility is achieved for values of $p_{\mathrm{charge}}$ above approximately 0.7. Consequently, $p_{\mathrm{charge}}=0.7$ is selected for the "Terminal only" strategy. Since the charging decisions are governed by a stochastic process, multiple realizations were performed for each tested value of $p_{\mathrm{charge}}$ to account for this variability and ensure the robustness of the calibration results.

\section{GTFS data pre-processing}
\label{app:gtfs_preprocessing}

The original GTFS feed provided by the CETUD contained explicit schedules for all individual trips but did not include a \texttt{frequencies.txt} file. Since GTFS4EV relies on a headway-based service representation, service frequencies were reconstructed from the scheduled trip data. Examination of the resulting patterns confirmed a high degree of operational regularity, making this reconstruction straightforward. Additional pre-processing included:

\begin{itemize}
\item Simplification of stop definitions by retaining only parent stations (stations in the original feed were modeled using separate directional platforms together with parent stations) and removing a small number of unused stops;
\item Removal of redundant and unused shapes in shapes.txt, retaining a single representative shape for the corridor;
\item Introduction of a 10 minute idle time at terminals following guidance from the the local transport authority, as idle times at terminals were not captured in the original GTFS feed.
\end{itemize}

The inspection of the GTFS data revealed two noteworthy operational characteristics: (1) the B2 schedule indicates that the first departure from Petersen occurs at 7:00 rather than 6:00, which is consistent with the bus schedules published online \cite{sunubrt}; and (2) a number of additional trips operate along the B1 line between Petersen and the Grand Médine stop during the morning and evening peak travel hours, thereby increasing the service frequency on this section of the line thanks to reinforcement buses. Hence, three distinct service patterns are observed: the B1 "Omnibus", the B2 "Semi-Express", and the B1 reinforcement buses operating during peak travel hours.

\section*{CRediT authorship contribution statement}
\label{authorship}
\textbf{Jérémy Dumoulin:} Conceptualization, Methodology, Software, Data curation, Investigation, Visualization, Writing - original draft, Writing -- review \& editing. \textbf{Cheikh Mouhamed Fadel Kebe:} Methodology, Resources, Investigation, Validation, Writing - review \& editing. \textbf{Babacar M. Ndiaye:} Resources, Validation, Writing - review \& editing. \textbf{Noémie Jeannin:} Methodology, Writing - review \& editing. \textbf{Christophe Ballif:} Supervision, Funding acquisition, Writing - review \& editing. \textbf{Nicolas Wyrsch:} Methodology, Supervision, Project administration, Funding acquisition, Writing - review \& editing.

\section*{Declaration of competing interest}
\label{competing-interest}
The authors declare no competing interests.

\section*{Acknowledgments}
\label{acknowledgments}
This work was supported by the HORIZON OpenMod4Africa project (Grant number 101118123), with funding from the European Union and the State Secretariat for Education, Research and Innovation (SERI) for the Swiss partners. The authors gratefully acknowledge the CETUD for providing the GTFS dataset used in this study and for their valuable feedback on the results.

\section*{Data availability}
\label{data}
The GTFS dataset is confidential and cannot be made publicly available. The GTFS4EV framework developed for this study is open-source, with detailed documentation for installation and usage available online \cite{gtfs4ev_docs} The analyses presented in this paper were performed using GTFS4EV version 0.2.4.

\section*{Declaration of AI-assisted technologies in the writing process}
\label{ai}
During the preparation of this work, the authors used IA tools to enhance its quality and clarity. After using these tools, the authors reviewed and edited the content as needed and take full responsibility for the content of the publication.


\bibliographystyle{elsarticle-num} 
\bibliography{references}

\end{document}